 \documentclass[12pt,preprint,epsfig]{aastex}

 \usepackage[]{graphicx}
 \usepackage{emulateapj5}

 \shorttitle{Cosmic Supernova Rates and the Hubble Sequence}
 \shortauthors{Calura \& Matteucci}

\begin{document}


 \title{Cosmic Supernova Rates and the Hubble Sequence}


 \author{Francesco Calura$^{1}$ and Francesca Matteucci$^{2,1}$}
\affil{
1 INAF - Osservatorio Astronomico di Trieste, via G.B. Tiepolo 11,  
34131 Trieste, Italy \\
2 Dipartimento di Astronomia-Universit\'a di Trieste, Via G. B. Tiepolo
  11, 34131 Trieste, Italy
}
 \email{fcalura@oats.inaf.it}




\begin{abstract}
We compute the type Ia, Ib/c and II supernova (SN) rates 
as functions of the cosmic time 
for galaxies of different morphological types. 
We use four different chemical evolution models, each one reproducing the features of 
a particular morphological type: 
E/S0, S0a/b, Sbc/d and Irr galaxies. We essentially describe the Hubble sequence by means of decreasing 
efficiency of star formation and increasing infall timescale. 
These models are used to study the evolution of the SN rates per unit luminosity and per unit mass as functions 
of cosmic time and as functions of the Hubble type. Our results indicate that: (i) the observed increase of the SN  
rate per unit luminosity 
and unit mass from early to late galaxy types is accounted for by our models. 
Our explanation of this effect is related to the fact that  
the latest Hubble types have the highest star formation rate per unit mass; 
(ii)  By adopting a Scalo (1986) initial mass function in spiral disks, we find that 
massive (i.e. with initial mass $>25 M_{\odot}$) single stars ending their lives as Wolf-Rayet objects
are not sufficient to account for 
the observed  type Ib/c SN rate per unit mass. Less massive stars (i.e. with initial masses $12 < M/M_{\odot} < 20$) 
in close binary systems can give instead a significant contribution to the local Ib/c SN rates. 
On the other hand, with the assumption of a Salpeter (1955) IMF for all galaxy types, single massive WR stars are sufficient to account for  
the observed  type Ib/c SN rate. 
(iii) our models allow us to reproduce the observed type Ia SN rate density up to redshift $z\sim 1$.  
At higher redshifts, our rates are higher than the few available data. In particular, we predict an increasing type Ia SN 
rate density with redshift, reaching a peak at redshift $z \ge3$, because of the contribution of massive spheroids.  
(iv) At $z=0$, we reproduce the observed CC SN rate density. Owing to the few available observations, no 
firm conclusion can be drawn on the behaviour of the CC SN rate at redshift $z>0$. 
\end{abstract} 




\section{Introduction}
The study of supernovae (SNe) plays a very important role in modern astronomy.  
SNe type Ia, Ib/c and II have different stellar progenitors, ending their lives on different timescales. 
Type Ib/c and II SNe have massive progenitors, exploding on short timescales, from 
few million to several tenths of million years, whereas 
Type Ia are believed to originate from low and intermediate mass stars, in particular from C-O white dwarfs 
in binary systems, ending their lives on 
timescales ranging from $\sim 0.03$ up to a Hubble time or more. 
The type Ia SN rate is hence related to the past star formation history of a galaxy, whereas the type Ib/c and II 
rates reflect  
the birth rate of massive stars, i.e. the galactic star formation rate (SFR).  
For this reason, the parallel study of the type Ia, Ib/c and II SNe provides us with a complete set 
of information about stellar and galactic evolution. \\
In the recent years, it has been possible to determine the local supernova rates (SNRs) in great detail, 
both as functions of 
the luminosity and mass of the various galactic morphological types, as well as in different environments 
(Muller et al. 1992, Cappellaro, Evans \& Turatto 1999, Gal-Yam, Maoz \& Sharon 2002, Mannucci et al. 2005).   
Furthermore, modern telescopes have allowed the astronomers to derive the supernova 
rates up to high redshift, providing crucial information on the various SN  progenitors and on the main 
 parameters determining galaxy evolution, such as the star formation rate and the stellar initial mass function.\\
This wealth of data has allowed us to improve our theoretical understanding of SNe and of their progenitors.
Several theoretical investigations of the cosmic SN rate  
have suggested some constraints on the cosmic star formation history (Madau, Della Valle \& Panagia 1998; Sadat et al.
 1998, Kobayashi, Tsujimoto \& Nomoto 2000, Hopkins \& Beacom 2004)  and on the timescales for the explosion of the 
progenitors of type Ia SNe (Yungelson \& Livio 2000, Mannucci et al. 2006). 
However, in all of these works, little emphasis has been put on the study of the SN rate as a function of the 
galactic morphological type. 
Another aspect which has not been investigated is how the SN rates in single galactic types have 
evolved with cosmic time, as well as the contribution to the cosmic SN rates 
by different galactic types.  
Furthermore, there has been no attempt to model the SN rate as a function of the 
galactic stellar mass. The study of these aspects is the main motivation of this paper. 
We use detailed chemical and spectrophotometric evolution models for galaxies of different morphological types 
and we attempt to study the Ia, Ib/c and II SNRs as functions of the galactic Hubble type 
and of the cosmic time. 
We build a galactic Hubble sequence, consisting of four classes of models: E/S0 
galaxies, spirals (S0a/b and Sbc/d) and irregular types (Irr). 
We model the SN rates for all known SN types (i.e. types Ia, Ib/c and II)  and by comparing our predictions with the observational data, 
we derive important information on the SN progenitors.
Finally, by studying the SN rate density and its evolution with redshift, we gain information on  
the global star formation rate and on how  
the galaxy populations evolved throughout a large fraction of the cosmic time.\\
The present paper is organized as follows. In section 2 we present the chemical evolution equations  
and the models used in this work. In section 3 we present our results. Finally, in section 4 we draw 
the conclusions.

\section{The chemical evolution models}

\subsection{How to model the galactic Hubble sequence}
In this paper, the chemical evolution 
models are used to describe four different galactic Hubble types: E/S0, S0a/b, Sbc/d and Irregulars. 
A model for a spheroid, characterized by a strong and rapid 
star formation event, followed by passive evolution, is used to describe 
E/S0 galaxies. 
Two different  models describe early type spirals (S0a/b) and late spirals  (Sbc/d), respectively.
Finally, to describe Irregular galaxies, 
we use a model characterized by a star formation history which has proceeded with bursts of low efficiency.\\
Here we present the basic chemical evolution equations and the assumptions common to all models. 

The time-evolution of the fractional mass of the element $i$ 
in the gas within a galaxy, $G_{i}$, is described by the basic 
equation:

\begin{equation}
\dot{G_{i}}=-\psi(t)X_{i}(t) + R_{i}(t) + (\dot{G_{i}})_{inf} -
(\dot{G_{i}})_{out}
\end{equation}

where $G_{i}(t)=M_{g}(t)X_{i}(t)/M_{tot}$ is the gas mass in 
the form of an element $i$ normalized to a total fixed mass 
$M_{tot}$ and $G(t)= M_{g}(t)/M_{tot}$ is the total fractional 
mass of gas present in the galaxy at the time $t$. 
The same quantities can be defined in terms of the surface gas and mass 
densities, especially in spiral galaxies. 
The quantity $X_{i}(t)=G_{i}(t)/G(t)$ represents the 
abundance by mass of an element $i$, with
the summation over all elements in the gas mixture being equal 
to unity. $\psi(t)$ is the fractional amount of gas turning into 
stars per unit time, namely the SFR. 
$R_{i}(t)$ represents the returned fraction of matter in the 
form of an element $i$ that the stars eject into the ISM through 
stellar winds and supernova explosions. This term contains all 
the prescriptions concerning the stellar yields and 
the SN progenitor models. \\
The two terms 
$(\dot{G_{i}})_{inf}$ and  $(\dot{G_{i}})_{out}$ account for 
the infall of external gas and for galactic winds, respectively.
The presence of infall and winds varies for galaxies of different 
morphological type (see sections 2.2, 2.3 and 2.4). \\
The nucleosynthesis prescriptions are common to all models. 
For massive stars and type Ia SNe, we adopt the empirical yields suggested by  Fran\c cois et al. (2004),   
which are substantially based on the Woosley \& Weaver (1995) and Iwamoto et al. (1999) yields, respectively. 
Fow low and intermediate mass stars, we adopt the prescriptions by van den Hoeck $\&$ Groenewegen (1997).  \\
The prescription adopted for the star formation history is the
main feature which characterizes a particular morphological 
galactic type.
In its simplest form, the SFR $\psi(t)$ in our models is a 
Schmidt (1959) law expressed as:

\begin{equation}
\psi(t) = \nu G^{k}(t)
\end{equation}

The quantity $\nu$ is the efficiency of star formation, 
namely the inverse of the typical time-scale for star formation,
and is expressed in $Gyr^{-1}$.\\
Unless otherwise stated, the rate of gas infall, for a given element i, is defined as:
\begin{equation}
(\dot G_{i})_{inf}\,=\, \frac{A}{M_{tot}}e^{-t/ \tau}
\end{equation}

with A being a suitable constant and $\tau$ the infall timescale.
The rate of gas loss via galactic winds for each element {\it i} is
assumed to be proportional to the star formation rate at the 
time {\it t}:

\begin{equation}
 \dot G_{iw}\,=\,w_{i} \, \psi(t)
\end{equation}

where $w_{i}$ is a 
free parameter describing the efficiency of the galactic
wind. 
In all models the instantaneous recycling approximation is relaxed and the 
stellar lifetimes are taken into account.

\subsection{Early type galaxies}
For the chemical evolution of ellipticals and S0 galaxies we adopt the model of Pipino \& 
Matteucci (2004) where we address the reader for details. 
Here we recall the main assumptions:
ellipticals form by means of a rapid collapse of pristine gas where star 
formation occurs at a very high rate (starburst-like regime), and after a 
timescale varying with the galactic mass, each galaxy develops a galactic wind 
due to the energy deposited by SNe into the interstellar medium (ISM). After 
the development of this wind, no star formation is assumed to take place.
The SN feedback is taken into account together with the cooling of SN 
remnants and the development of galactic winds is calculated in a 
self-consistent way. Massive but diffuse haloes of dark matter around these 
galaxies are considered.
We assume that the efficiency of star formation rate is higher in more 
massive objects which evolve faster than less massive ones (inverse-wind 
scenario, Matteucci, 1994, otherwise called ``downsizing'').
A Salpeter (1955) IMF constant in space and time is adopted.
 The choice of such an IMF for ellipticals and S0 galaxies assures that 
several observational constraints such as the average stellar abundances and the color-magnitude diagram 
(see Pipino \& Matteucci, 2004) are well reproduced, as well as the metal content in clusters of galaxies (see Renzini 2004).  
In the SFR expression (eq. 2) we assume k=1 and $\nu =10 Gyr^{-1}$. 
The infall is assumed to occur on an extremely short timescale ($< 0.1$ Gyr).

\subsection{Spiral galaxies} 

The model used for spiral (S0a/b and Sbc/d) galaxies is the two infall model 
by Chiappini et al. (1997). 
The first infall creates the halo and the thick 
disk, on a timescale of $\sim 1$ Gyr.  The second infall gives rise to the thin disk. 
The timescale for the 
disk formation is assumed to increase 
with the galactocentric distance ($\tau= 8$ Gyr at the solar circle), 
thus producing an ``inside-out'' scenario 
for the disk formation. The use of the two-infall model to describe spirals other than the Milky Way is 
motivated by the fact that the Milky Way is a rather common spiral galaxy. Recent results of Yoachim \& Dalcanton (2006), who 
have analyzed a sample of nearby disks,  
show that the structural properties of the thin and thick disks of these galaxies are similar 
to the ones of the Milky Way. This indicates that all spirals are likely to have a similar origin. 

The infall law used for spirals is based on  equation (3) but is expressed in terms of the 
surface mass density. 
For a generic chemical element $i$, the accretion term in spirals is given by: 

\begin{equation}
(\dot{G_{i}})_{inf}= X_{i, inf}(A e^{-t/\tau_H}+Be^{-(t-\tau_{inf})/ \tau_D})
\end{equation}
with $B=0$ for $t<\tau_{inf}$.
The quantity $\tau_H$ is the timescale for the inner halo formation (0.5-1 Gyr)
and  $\tau_D$ is the thin disk timescale, whereas 
$\tau_{inf}$ is the time of maximum gas accretion onto the disk, coincident with the end of the halo/thick 
disk phase. Both $\tau_D$ and $\tau_{inf}$ assume different values for 
the S0a/b and Sbc/d types (see table~\ref{parameters}). 
$X_{i, inf}$ is the abundance of the element $i$ in the infalling material, that we assume to have a 
primordial chemical composition. 
The quantities $A$ and $B$ are constants derived in order to reproduce 
the current average total surface mass density in the halo and in the disk of each galactic type, respectively. 
For the surface mass density in the halo, the reference value is the one of the Milky Way, 
i.e. $\sigma_{H}=20 \,  M_{\odot}\, pc^{-2}$ (Kuijen \& Gilmore 1991). For the surface mass density in the disks, 
the values we adopt are: 
the total surface mass densities by Roberts \& Haynes (1994), i.e.  $\sigma_{D, S0ab}= 154 M_{\odot}\, pc^{-2} $ 
and $\sigma_{D, Sbcd} \sim 100 M_{\odot}\, pc^{-2}$  for S0a/b and Sbc/d galaxies, respectively.

The IMF is assumed to be constant in space and time.
 For the IMF, we test two possibililities: the Salpeter (1955) IMF and the Scalo (1986) IMF. 
 In general, for spiral disks the latter is to be preferred over the former, for several reasons. 
With this IMF, in fact,  it is possible to explain  
the I-band mass to light ratio for the stellar component of spiral galaxies, whereas a Salpeter 
IMF leads to an overestimation of this quantity (Portinari et al. 2004). 
An important indication about the disk IMF comes also from chemical evolution models (see Chiappini et al, 1997; 2001), which clearly
indicate that to reproduce the main features of the solar neighbourhood and the whole disk a Scalo-like IMF is 
preferred, and that the Salpeter IMF would overestimate the solar abundances (see Romano et al. 2004 for a detailed discussion on this point).

\subsection{Dwarf irregulars}
For the dwarf irregulars we assume a
one zone model with instantaneous and complete mixing of gas inside
this zone. Irregular galaxies 
assemble all their mass by means of a continuous infall 
of pristine gas.
The SFR is characterized by bursts separated by
long quiescent periods. 
These star bursts, if strong enough, can also trigger a galactic wind (GW) (Bradamante et al. 1998, Recchi et al. 2002).
The conditions for the onset of the GW are identical to the ones described for elliptical galaxies. 
As for E/S0 galaxies, for Irr the adopted IMF is the Salpeter (1955) one. 
This choice is in agreement with the results by Calura \& Matteucci (2004), who have shown that the 
adoption of a steeper IMF in dwarf irregulars leads to an underestimation of their average metallicity. 
Here we adopt a star formation history suitable for the Large Magellanic Cloud as 
representative of this type of galaxies (Calura, Matteucci \& Vladilo 2003).\\
The spectral evolution of all the galactic morphological types has been calculated by means 
of the spectro-photometric code by Jimenez et al. (2004). This code  allows one 
to follow in detail the metallicity evolution of the gas out of which the stars form. 
This is possible thanks to the large number of simple stellar populations calculated by Jimenez et al. (2004) by means of new stellar tracks, 
with ages between $1 \times 10^{6}$ and $14 \times 10^{9}$ yr and metallicities ranging from $Z=0.0002$ to $Z=0.1$. \\
The main spectral observables considered in this 
work (i.e. the galactic blue luminosities) have been  corrected for dust extinction effects (Cappellaro et al. 1999). 
For this reason, in our spectro-photometric calculations we do not take into account dust extinction.\\

In table~\ref{parameters} we show the adopted parameters for the four chemical evolution models described in this section.  
In column 1 we list the parameters which are, for all galaxies, 
the star formation efficiency $\nu$, the infall timescale $\tau_{inf}$ and the IMF.  
For the spirals, there is one additional parameter, i.e. the thin disk timescale $\tau_D$. 
In columns 2, 3, 4 and 5 we present the parameter values adopted for E/S0, S0a/b, Sbc/d and Irr, respectively. 
The SFR parameters adopted in this work have been chosen in order to reproduce the main features 
of the Hubble sequence, i.e. the fact that galaxies from E/S0 to Irr show progressively  younger stellar 
populations, higher star formation rates per unit mass and higher gas fractions (Sandage 1986, Kennicutt 1998).  
 In particular, we interpret the Hubble sequence as a sequence of decreasing efficiency of SF and increasing infall 
timescale. \\
In table~\ref{properties}, we show the present-day main properties of the galactic morphological types studied in this work, 
as observed by various authors and as predicted by means of our models. We assume that the age of all the galaxies 
is 12.5 Gyr, which, assuming the set of cosmological parameters suggested by Spergel et al. (2006) ($\Omega_{0}=0.24$, 
 $\Omega_{\Lambda}=0.76$, $h=0.73$), corresponds to a redshift of formation $z_{f}=5$. The observables considered 
in table~\ref{properties} are the neutral H mass M$_{HI}$, the blue luminosity L$_{B}$, 
the ratio between these two quantities, the (B-V) colour and the metallicity. \\
The measure of the metallicty in the ISM of spirals and irregular galaxies is possible by means of the observation of the 
bright H$_{II}$ regions, whereas the determination of the abundances in the hot ISM of the elliptical and S0 galaxies can be a very diffucult task. Moreover, the hot gas in ellipticals may have an external origin.
The abundances in E/S0 galaxies are therefore stellar abundances. They are derived from their visual integrated spectrum by means of metallicity indices and then converted to [Fe/H] through suitable calibrations (see Kobayashi \& Arimoto 1999). 
From table~\ref{properties}, 
it is possible to see that all the properties of present-day galaxies are reproduced with good accuracy by our models.

\subsection{The star formation rates}
In figure 1, we plot the time evolution of the star formation rates (expressed in $M_{\odot}/yr$) 
for the four models described above. A typical 
E/S0 galaxies is characterized by very high SFR values (from 100 to 1000  $M_{\odot}/yr$) and by a 
starburst lasting $\sim 0.2$ Gyr. \\
The predicted SFR  for S0a/b and Sbc/d  are characterized by two peaks, which are 
due to the two infall episodes.
At the present time the S0a/b and Sbc/d models have similar SFR values, of the order of  
$\sim 4-6 M_{\odot}/yr$ but in the past the S0a/b had a higher SFR than the Sbc/d  during the disk phase. 
In fact, the timescale for gas accretion onto the disk in Sbc/d galaxies is assumed to be larger than 
in S0a/b, thus producing a shallower SFR. 
The SFR of the irregular model consists of an early burst, lasting 0.1 Gyr, and 
a very long period of low efficiency star formation, lasting 9 Gyr. 
The irregular model is the one characterized by the lowest SFR values, typically of the order of $\sim 0.1 - 0.2 M_{\odot}/yr$.\\
In Figure 2, we show the predicted evolution of the SFR per unit mass ($SFR_{m}$, expressed in $10^{9} \,  yr^{-1}$) 
for the four models used in this work. This quantity was introduced by Sandage (1986) and is very 
important for the arguments discussed in this paper. In fact, this quantity is the main driver of  
the morphological differences among the various Hubble types. 
From Figure ~\ref{SFRuM}, we see that the galaxy types presenting the highest $SFR_{m}$ are the E/S0, 
with values up to several $10^{-7} \,  yr^{-1}$. At the beginning of the bursts, also Irr galaxies 
present a very high $SFR_{m}$ of $\sim 10^{-7} \,  yr^{-1}$. 
S0a/b and Sbc/d have initial  $SFR_{m}$  of $\sim  10^{-7} \,  yr^{-1}$ only during 
a very short initial phase, after which they evolve with  progressively decreasing values. 
It is interesting to note that at the present time, the $SFR_{m}$ is an increasing function of the 
Hubble type, with progressively higher values from the early to the late types. This fact will have 
important consequences on the study of the SNR per unit luminosity and mass (see sections 3.1 and 3.2). 

\renewcommand{\baselinestretch}{1.0}
\begin{table*}
\centering
\begin{tabular}{lccccc}
\\[-2.0ex] 
\hline
\\[-2.5ex]
\multicolumn{1}{l}{}&\multicolumn{2}{c}{}&\multicolumn{1}{c}{Parameters}&\multicolumn{2}{c}{}\\
\hline
\multicolumn{1}{c}{Hubble Type}&\multicolumn{1}{c}{}&\multicolumn{1}{c}{$\nu$}&\multicolumn{1}{c}{$\tau_{inf}$}&\multicolumn{1}{c}{$\tau_D$}&\multicolumn{1}{c}{IMF}\\
\multicolumn{1}{c}{}&\multicolumn{1}{c}{}&\multicolumn{1}{c}{(Gyr$^{-1}$)}&\multicolumn{1}{c}{(Gyr)}&\multicolumn{1}{c}{(Gyr)}&\multicolumn{1}{c}{}\\
\hline
\hline
\\[-1.0ex]
E/S0                &       &   10     &    -     &      -        &   Salpeter     \\
S0a/b               &       &    2     &    1     &      3        &   Scalo        \\
S0a/b               &       &    2     &    1     &      1        &   Salpeter     \\
Sbc/d               &       &   0.8    &    5     &      5        &   Scalo        \\
Sbc/d               &       &   0.8    &    5     &      5        &   Salpeter     \\
Irr                 &       &   0.05   &    10    &      -        &   Salpeter     \\
\hline
\hline
\end{tabular}
\label{parameters}
\caption{}
Adopted parameters for the galactic Hubble types modeled in this work. 
The galactic Hubble types  are listed in column 1. In  
columns 2, 3, 4 and 5, we present the adopted parameters, i.e.  the star formation efficiency $\nu$, the infall timescale $\tau_{inf}$, 
the disk timescale $\tau_D$ and the IMF, respectively.
\end{table*}
\renewcommand{\baselinestretch}{1.0}
\begin{table*}
\centering
\caption{}
\begin{tabular}{lcccccccccccccc}
\\[-2.0ex] 
\hline
\\[-2.5ex]

\multicolumn{1}{l}{Hubble Type}&\multicolumn{2}{c}{L$_{B}$}&\multicolumn{2}{c}{M$_{HI}$}&\multicolumn{2}{c}{M$_{HI}$/L$_{B}$}&\multicolumn{2}{c}{B-V}&\multicolumn{2}{c}{Metallicity}\\
\multicolumn{1}{l}{}&\multicolumn{2}{c}{(10$^{10}$L$_{\odot}$) }&\multicolumn{2}{c}{(10$^{8}$M$_{\odot}$)}&\multicolumn{2}{c}{(M$_{\odot}$/L$_{\odot}$)}&\multicolumn{2}{c}{}&\multicolumn{2}{c}{}\\
\hline 
\multicolumn{1}{c}{}&\multicolumn{1}{c}{Obs}&\multicolumn{1}{c}{Pred}&\multicolumn{1}{c}{Obs}&\multicolumn{1}{c}{Pred}&\multicolumn{1}{c}{Obs}&\multicolumn{1}{c}{Pred}&\multicolumn{1}{c}{Obs}&\multicolumn{1}{c}{Pred}&\multicolumn{1}{c}{Obs}&\multicolumn{1}{c}{Pred}\\

\hline 
\hline
\\[-1.0ex]
\multicolumn{1}{l}{E/S0}&\multicolumn{1}{c}{0.3-9.4$^1$}&\multicolumn{1}{c}{0.9}&\multicolumn{1}{c}{0.04-50$^1$}&\multicolumn{1}{c}{0.07}&
\multicolumn{1}{c}{0.0001-0.13$^1$}&\multicolumn{1}{c}{0.0008}&\multicolumn{1}{c}{0.86-0.94$^2$} &\multicolumn{1}{c}{0.86} &
\multicolumn{2}{c}{$<$[Fe/H]$>$$^a$$^{5}$=}\\
       &   &      &     &      &    &    &    &  &  -0.8-0.3    &   -0.69 \\ 
\hline
\\[-1.0ex]
\multicolumn{1}{l}{S0a/b}&\multicolumn{1}{c}{2.1-10.7$^2$}&\multicolumn{1}{c}{4.3}&\multicolumn{1}{c}{18-260$^2$}&\multicolumn{1}{c}{65}&
\multicolumn{1}{c}{0.04-0.33$^2$}&\multicolumn{1}{c}{0.15}&\multicolumn{1}{c}{0.55-0.83$^2$}&\multicolumn{1}{c}{0.66} &
\multicolumn{2}{c}{12+log(O/H)$^a$$^{6}$=}\\
(Scalo IMF)   &   &      &     &      &    &    &    &  &  8.0-9.5    &   8.8 \\ 
\multicolumn{1}{l}{S0a/b}&\multicolumn{1}{c}{2.1-10.7$^2$}&\multicolumn{1}{c}{2.9}&\multicolumn{1}{c}{18-260$^2$}&\multicolumn{1}{c}{47.5}&
\multicolumn{1}{c}{0.04-0.33$^2$}&\multicolumn{1}{c}{0.16}&\multicolumn{1}{c}{0.55-0.83$^2$}&\multicolumn{1}{c}{0.69} &
\multicolumn{2}{c}{12+log(O/H)$^a$$^{6}$=}\\
(Salp. IMF)   &   &      &     &      &    &    &    &  &  8.0-9.5    &   9.3 \\ 
\hline
\\[-1.0ex]
\multicolumn{1}{l}{Sbc/d}&\multicolumn{1}{c}{0.98-9.55$^2$}&\multicolumn{1}{c}{4.7}&\multicolumn{1}{c}{40-260$^2$}&\multicolumn{1}{c}{125}&
\multicolumn{1}{c}{0.19-0.56$^2$}&\multicolumn{1}{c}{0.27}&\multicolumn{1}{c}{0.42-0.62$^2$}&\multicolumn{1}{c}{0.48} &
\multicolumn{2}{c}{12+log(O/H)$^a$$^{6}$=}\\
(Scalo IMF)   &   &      &     &      &    &    &    &  &  8.0-9.5    &   8.6 \\ 
\multicolumn{1}{l}{Sbc/d}&\multicolumn{1}{c}{0.98-9.55$^2$}&\multicolumn{1}{c}{4.7}&\multicolumn{1}{c}{40-260$^2$}&\multicolumn{1}{c}{133}&
\multicolumn{1}{c}{0.19-0.56$^2$}&\multicolumn{1}{c}{0.28}&\multicolumn{1}{c}{0.42-0.62$^2$}&\multicolumn{1}{c}{0.44} &
\multicolumn{2}{c}{12+log(O/H)$^a$$^{6}$=}\\
(Salp. IMF)   &   &      &     &      &    &    &    &  &  8.0-9.5    &   9.0 \\ 
\hline
\\[-1.0ex]
\multicolumn{1}{l}{Irr}&\multicolumn{1}{c}{0.1-0.7$^2$}&\multicolumn{1}{c}{0.21}&\multicolumn{1}{c}{7.4-61.7$^2$}&\multicolumn{1}{c}{36.5}&
\multicolumn{1}{c}{0.36-2$^{3,4}$}&\multicolumn{1}{c}{1.7}&\multicolumn{1}{c}{0.35-0.53$^2$}&\multicolumn{1}{c}{0.38} &
\multicolumn{2}{c}{12+log(O/H)$^a$$^{6}$=}\\
(Scalo IMF)   &   &      &     &      &    &    &    &  &  7.5-9.0    &   8.2 \\ 

\hline
\hline
\end{tabular}
\flushleft
Observed and predicted present-day properties of the galactic morphological types studied in this work. 
The galactic Hubble types  are listed in column 1. In  
columns 2, 3, 4, 5 and 6  we present the observed and predicted values for the blue luminosity, the HI mass, the HI mass to light ratio, 
the B-V colour and the metallicity, respectively. 

\emph{References}: $^1$Sansom et al. (2000); $^2$Robert \& Haynes 1994, $^3$Garland et al. (2004); $^4$Hunter \& Elmegreen (2004); $^{5}$Kobayashi \& Arimoto (1999); $^{6}$Vila-Costas \& Edmunds (1992). \\
\emph{Notes}:$^a$For E/S0 galaxies, the chemical abundances reported in the table are the stellar ones. For S0a/b, Sbc/d and Irr galaxies, the chemical 
abundances reported in the table are the ones measured in H$_{II}$ regions.
\label{properties}
\end{table*}
\begin{figure*}
\centerline{\includegraphics[height=19pc,width=19pc]{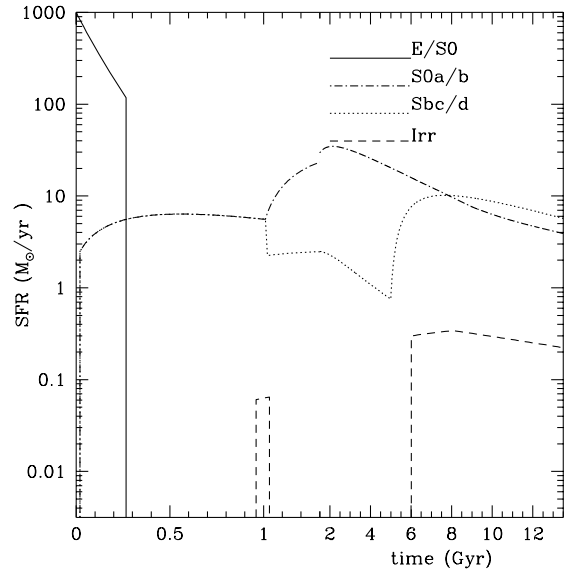}} 
\caption[]{Predicted star formation rate, expressed in $M_{\odot}/yr$, as a function of time for different galactic Hubble types. 
\label{SFR}
}
\end{figure*}

\begin{figure*}
\centerline{\includegraphics[height=19pc,width=19pc]{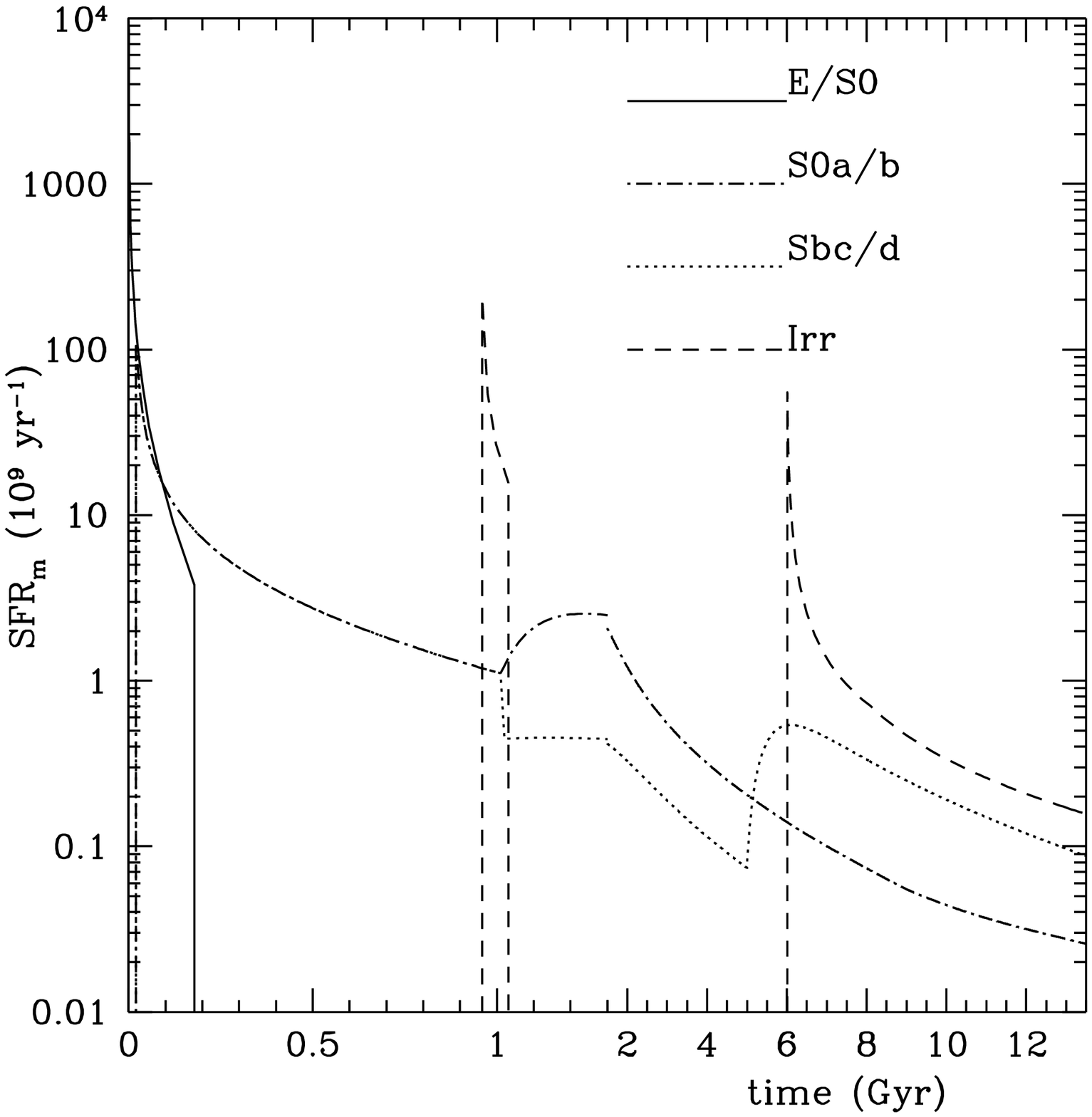}} 
\caption[]{Predicted star formation rate per unit mass, expressed in $ 10^{9} \cdot yr^{-1}$, as a function of time for different galactic Hubble types. 
\label{SFRuM}
}
\end{figure*}

\subsection{The supernova Rates}

\subsubsection{Type Ia supernovae}

To describe type Ia supernova progenitors, we assume the single degenerate (SD) scenario by Whelan \& Iben (1973). 
In this scenario, a C-O white dwarf accretes mass
from a non-degenerate companion until it reaches the Chandrasekhar mass ($\sim 1.4 M_{\odot}$) 
and explodes via C-deflagration, leaving no remnant.\\
The type Ia SN rate is expressed as:
\begin{equation}
 R_{Ia}(t) = A_{Ia} \int_{M_{Bm}}^{M_{BM}} \phi(M_{B}) [\int_{\mu_{m}}^{0.5} f(\mu)
\psi(t-\tau_{M_{2}})d\mu]dM_{B}
\end{equation}
where $A_{Ia}$ represents the proportion of stars in the mass range $M_{Bm}\le M_{B} \le M_{BM}$, which are born as binaries of that particular type which eventually produces SNe Ia. 
The quantity $\mu=M_{2}/M_{B}$ is the ratio between the secondary component of the binary system (i.e. the less
massive one) and the total mass of the system and $f(\mu)$ is the 
distribution function of this ratio. Statistical studies indicate that mass 
ratios close to $0.5$ are preferred, so the formula:\\
\begin{equation}
 f(\mu)=2^{1+\beta}(1+\beta)\mu^{\beta}
\end{equation}
is commonly adopted (Matteucci \& Recchi 2001), with $\beta=2$ as a parameter.
$\tau_{M_{2}}$ is the lifetime of the secondary star in the system, 
which determines the time-scale for the explosion.
The assumed value of $A_{Ia}$ is fixed by reproducing the present time observed rate and depends on the assumed IMF 
(Matteucci et al. 2003). 
For the masses ${M_{Bm}}$ and ${M_{BM}}$, we
chose the values $3M_{\odot}$ and $16M_{\odot}$, respectively (see Matteucci \& Greggio 1986).\\


\subsubsection{Type II supernovae}
We assume that single massive stars with initial masses in the range $>8 - 25M_{\odot}$ explode as type II core collapse 
supernovae. 
The type II SN rate is expressed as: 

\begin{math}
R_{II}(t) = (1-A_{Ia}) \int_{>8}^{M_{BM}} \phi(M) \psi(t-\tau_{M})dM + 
\end{math}
\begin{equation}
 \int_{M_{BM}}^{25} \phi(M) \psi(t-\tau_{M})dM 
\end{equation}
The upper limit for for type II SN is uncertain and it can vary from 25 $M_{\odot}$ to 100 $M_{\odot}$, according to the mass loss.

\subsubsection{Type Ib/c supernovae}
The origin of type Ib/c SNe is rather controversial. 
They occur in late type galaxies, in general in the vicinity of star forming regions 
(Filippenko 1991).  This fact indicates that the most likely progenitors of these SNe are massive 
stars, such as Wolf-Rayet (WR) stars (Filippenko \& Sargent 1986, Schaeffer, Cass\'e \& Cahen 1987), which eject their 
H envelope by means of intense stellar 
winds. However, there are observational evidences against this hypothesis, such as the observed light curves, in general 
broader than the ones achievable by assuming that they originate from the collapse of very massive stars. Other problems are 
represented by the apparent paucity of WRs, unable to account for the rates observed in local galaxies (Muller et al. 1992), 
along with the fact that some events have been observed  relatively far from active star forming regions (Filippenko 1991). 
More promising candidates, which could account for these evidences, are represented by less massive stars, 
with typical masses of $12 - 20M_{\odot}$ (Baron 1992, Pols \& Nomoto 1997),  
in close binary systems. In this case, the loss of  their envelope occurs by means of Roche-Lobe 
overflows. However, we will start with the simplest assumption and, 
as suggested by Maeder (1992), we assume that all the massive stars with  initial mass $M \ge 25 M_{\odot}$ 
explode as type Ib/c supernovae. The Ib/c SN rate is then expressed as:

\begin{equation}
R_{Ib/c}(t) =  \int_{25}^{100} \phi(M) \psi(t-\tau_{M})dM 
\end{equation}

\section{Results:supernova rates per unit luminosity and per unit mass}

The SNR can be expressed per unit blue luminosity, according to:\\
\begin{equation}
 1 SNu=1SN/10^{10}L_{\odot B}/century
\end{equation}
The SNR per unit luminosity (SNu) depends on the details of the SN progenitor models and on the blue luminosity. 
On the other hand, the  SNR per unit stellar mass (SNuM, Mannucci et al. 2005) is expressed as:
\begin{equation}
 1 SNuM=1SN/10^{10}M_{\odot}/century
\end{equation} 
By adopting these units, the SNR contains information on the stellar mass, which reflects the 
integrated star formation history, hence it provides different pieces of information than the SNR per unit luminosity.

\subsection{SN rates per unit blue luminosity}
\begin{figure*}
\centerline{\includegraphics[height=45pc,width=45pc]{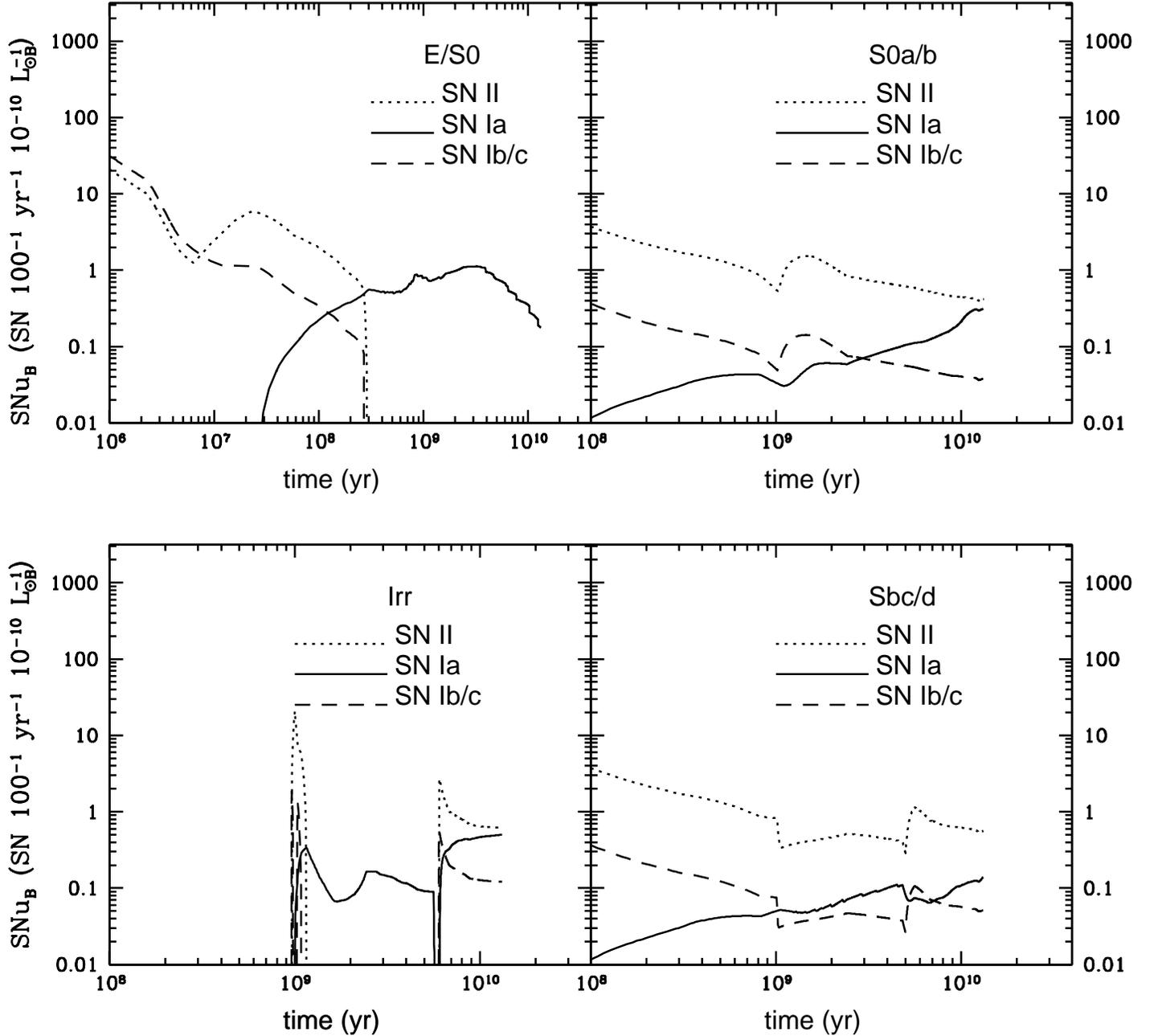}} 
\caption[]{Predicted time evolution of the type Ia (solid lines), type Ib/c (dashed lines) and type II (dotted lines) 
supernova rates per unit B luminosity (SNu) for galaxies of different Hubble types, as indicated in each figure, assuming a Salpeter (1955) 
IMF in E/S0 and Irr galaxies and a Scalo (1986) IMF in S0a/b and Sbc/d galaxies. }
\label{SNu}
\end{figure*}

In Figure ~\ref{SNu}, we show the predicted evolution of the type Ia, Ib/c and II SNRs 
per unit blue luminosity 
for all the 4 galactic morphological types studied in this work.  
In E/S0 galaxies, all type II SNe explode until $\sim 0.2$ Gyr, which corresponds to the  starburst 
timescale.  The Ib/c SNR is dominant at the beginning of the starburst, when the most massive stars 
explode. After  $\sim 0.01$Gyr, the type II SN dominate the total rate until $\sim 0.2$ Gyr. 
The type Ia SNR, on the other hand, starts to be significative at $\sim 0.1$ Gyr and peaks at $\sim 3$ Gyr. 
After this time, it decreases continuously up to the present time ($T_{0} \sim 12.5$ Gyr).\\  
The Ib/c and II SNRs of the S0a/b type reflect the effects of the two infall episodes, 
which determine the shape of the star formation history. 
They both  decrease until 1 Gyr,  
where they rise again and present a peak.  This peak 
occurs in correspondence of the discontinuity in the SFR at $t=1$ Gyr (see Fig. ~\ref{SFR}). 
After this peak, 
the Ib/c and II SNRs decrease continuously until the present time (12.5 Gyr). \\
The type Ia rate has a completely different behaviour. It presents a very broad peak 
in correspondence of the first infall, then it increases  nearly  
monotonically  until the present time. \\
Also for galaxies of the Sbc/d types, 
the Ib/c and II SNR strongly depend on the two infalls. 
They both decrease 
until 1 Gyr, when the SFR has a strong discontinuity (see Fig.~\ref{SFR}). 
The Ib/c and II SNRs are then nearly constant until 6 Gyr, when they present a peak and then decrease 
 up to the present time. 
The type Ia SNR is similar to the one described for the S0a/b galactic type.\\
For the Irr galaxy types, the SNRs strongly reflect the bursty star formation history. 
The Ib/c and II SNRs peak in correspondence of the first starburst, 
at 1 Gyr, then they drop as a consequence of the short duration of the first 
burst, which is 0.1 Gyr. The Ib/c and II SNRs present a second peak at $t\sim 6 $ Gyr, 
when the star formation starts again, then they decline until the present time. 
The type Ia SNR shows a first peak, occurring at 1.1 Gyr. A second broad peak is 
located at 3 Gyr. This second peak is mainly due to a strong decrease of the blue luminosity, 
predicted by our spectro-photometric model in the period in-between the two bursts. 
The type Ia SNR restarts to increase a short time after the second 
burst and then flattens, to remain constant up to 12.5 Gyr.\\

\begin{figure*}
\centerline{\includegraphics[height=45pc,width=45pc]{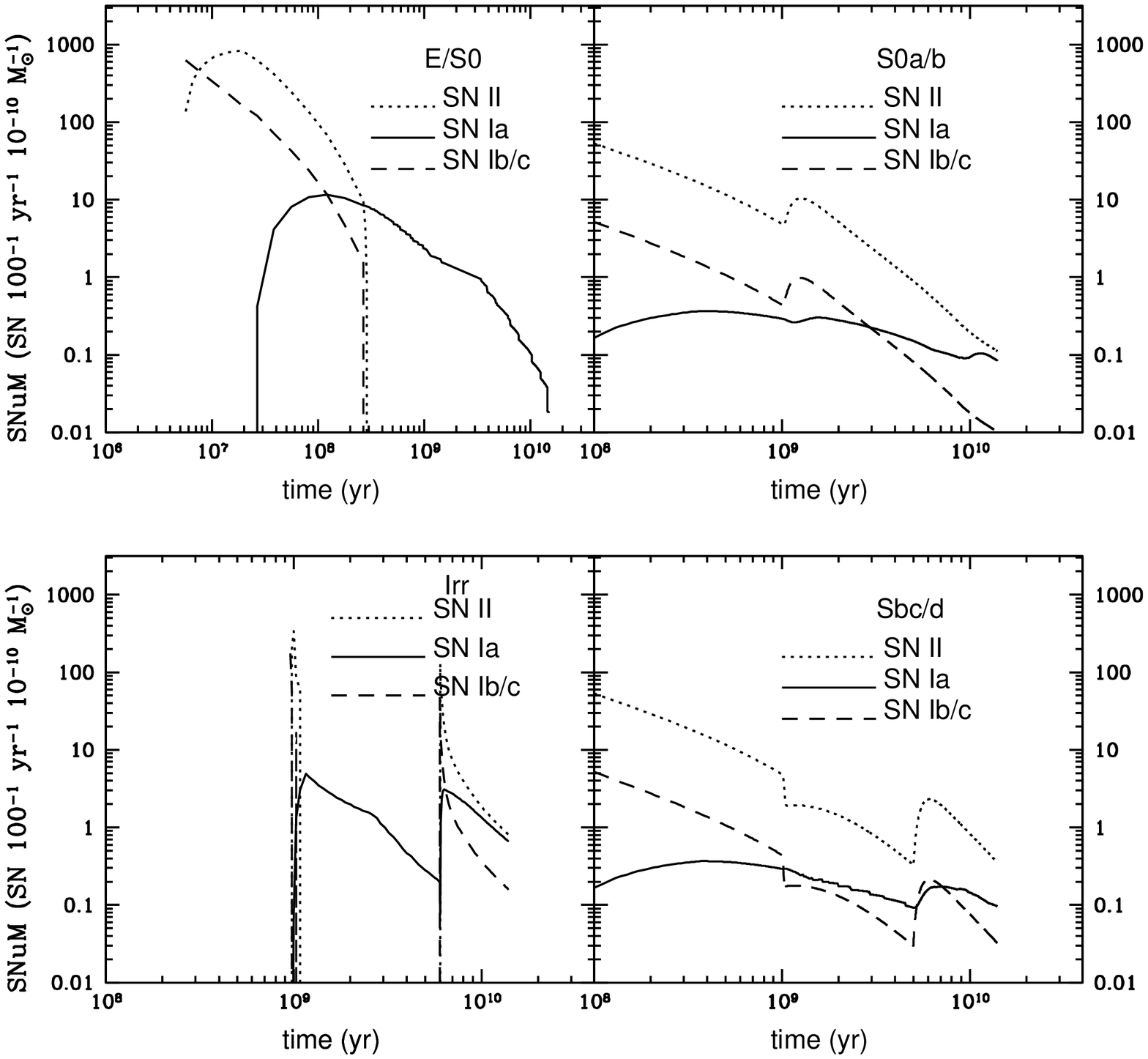}} 
\caption[]{Predicted time evolution of the type Ia (solid lines), type Ib/c (dashed lines) and type II (dotted lines) 
supernova rates per unit mass (SNuM) for galaxies of different Hubble types, as indicated in each figure, 
assuming a Salpeter (1955) 
IMF in E/S0 and Irr galaxies and a Scalo (1986) IMF in S0a/b and Sbc/d galaxies. }
\label{SNuM}
\end{figure*}

The predicted 
evolution of the SNR per unit mass for different galactic morphological types 
is shown in Figure~\ref{SNuM}. For each galactic type, the time evolution of the  Ib/c and II  SNRs are 
very similar to the SNRs per unit luminosity, as described above. 
However, we note that the predicted evolution of the type Ia SNR per unit mass  
is substantially different than the predicted type Ia SNR per unit luminosity. 
In the case of E/S0 galaxies, the SNuM peaks at a different time than the SNu. 
In the case of the S0a/b and Sbc/d galaxies, in the disk phase, i.e. at times larger 
than 1 Gyr, the SNuM is a decreasing function of time, whereas the SNu follows an opposite trend. 
Also for Irr galaxies, during the second star formation episode, the SNuM decreases, whereas the SNu 
increases. All of these effects are due to the fact that, as the cosmic time increases, 
the stellar mass and the  blue luminosity are increasing and decreasing, respectively.


\begin{figure*}
\centerline{\includegraphics[height=19pc,width=19pc]{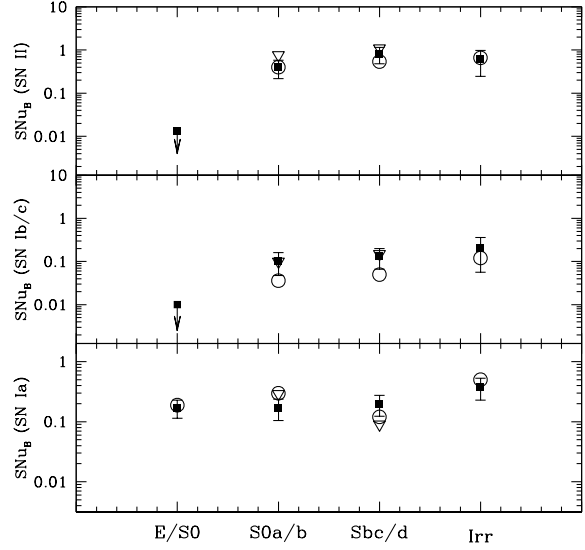}} 
\caption[]{
Observed and predicted present-day  Ia (lower panel), Ib/c (middle panel) and II (higher panel)  
SN rate per unit blue luminosity as a function of the galactic morphological type. 
The solid squares with the error bars are the observed values by Cappellaro et al. (1999).

The open circles represent the predicted values for the various galactic types, assuming 
that the progenitors of SNe Ib/c are single stars with masses $M \ge 25 M_{\odot}$, losing their envelope as Wolf-Rayet 
(WR) stars. The open circles are calculated  assuming a Salpeter IMF (1955) in E/S0 and Irr galaxies 
and a Scalo (1986) IMF in S0a/b and Sbc/d galaxies. 
 The inverted triangles are the SN rates calculated for S0a/b and Sbc/d galaxies, assuming a Salpeter (1955) IMF. 
 }
\label{SNu_}
\end{figure*}

\begin{figure*}
\centerline{\includegraphics[height=19pc,width=19pc]{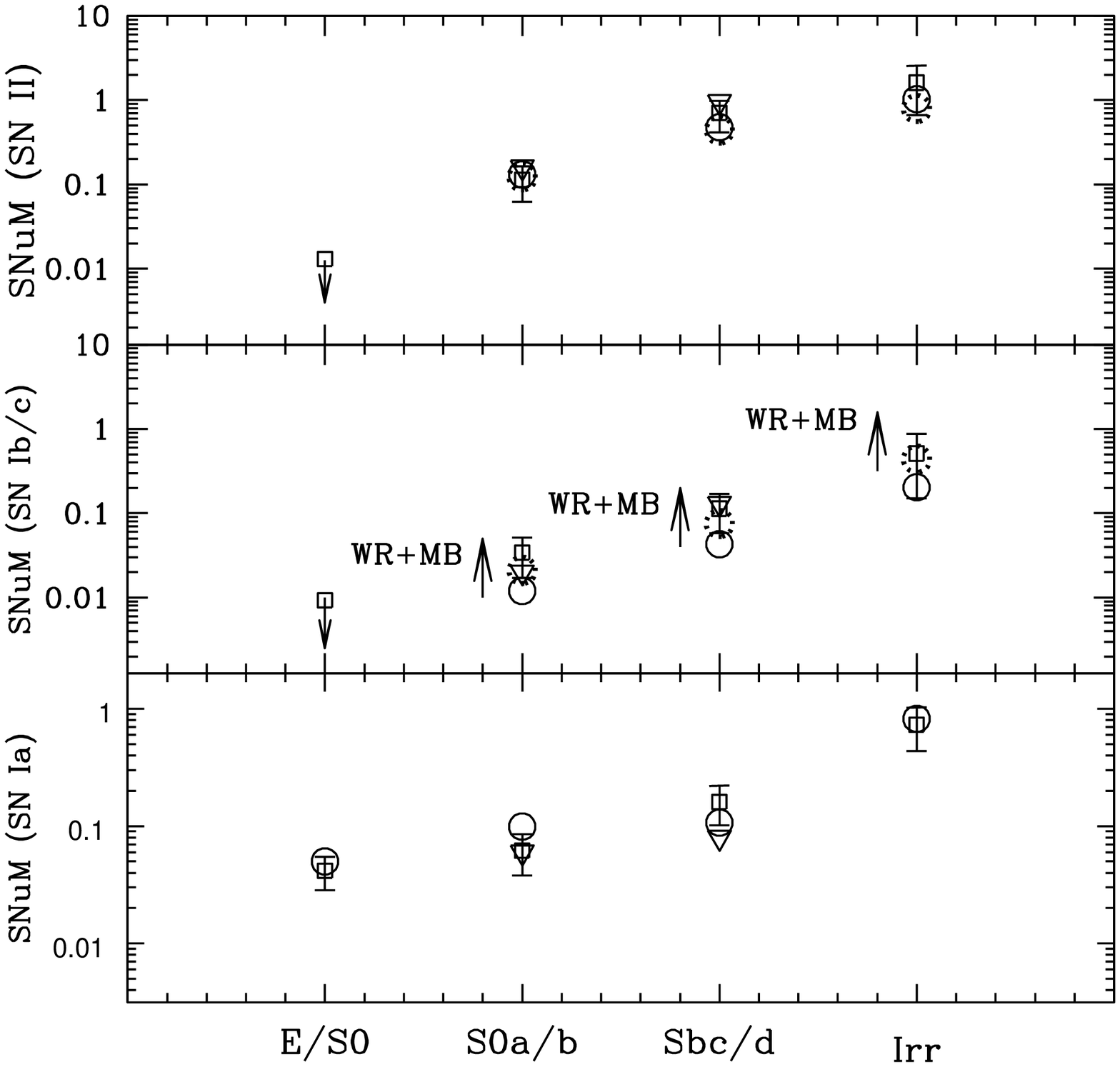}} 
\caption[]{Observed and predicted present-day  Ia (lower panel), Ib/c (middle panel) and II (higher panel)  
SN rate per unit mass as a function of the galactic morphological type. 
The open squares with the error bars are the observed values by  Mannucci et al. (2005). 
The open circles represent the predicted values for the various galactic types, assuming 
that the progenitors of SNe Ib/c are single stars with masses $M \ge 25 M_{\odot}$, losing their envelope as Wolf-Rayet 
(WR) stars. 
The dotted circles are the predicted values assuming 
that the progenitors of SNe Ib/c are WR stars plus 
massive binaries (MB). 
The open circles and the dotted circles are calculated  assuming a Salpeter (1955) IMF in E/S0 and Irr galaxies 
and a Scalo (1986) IMF in S0a/b and Sbc/d galaxies. 
 The inverted triangles are the SN rates calculated for S0a/b and Sbc/d galaxies, assuming a Salpeter (1955) IMF and that 
that the progenitors of SNe Ib/c are single stars with masses $M \ge 25 M_{\odot}$. 
}
\label{SNuM_}
\end{figure*}

In Figure~\ref{SNu_}, we plot the observed and predicted present-day  type 
Ia (lower panel), Ib/c (middle panel) and II (higher panel)  
SN rate per unit blue luminosity as a function of the galactic morphological type. 
We assume that all galaxies are coeval, with ages of 12.5 Gyr. 
The open circles represent the predictions, whereas the observations by Cappellaro et al. (1999) 
are represented by the solid squares. For all SN types, 
the observations indicate an increase of the SNu from early to late 
types. The interpretation of this trend is related to the fact that 
the late types have in general high SFR values per unit mass (see Figure 2), hence higher SN rates. 
Furthermore, the blue luminosity, in absence of active star formation,  is proportional to the stellar mass, which is large in 
early galaxy types (E/S0)  and small in late types, such as Sbc/d and Irr. 
The increasing trend of the observed SNu towards later Hubble types is due to 
a combination of both these effects. 
However, it is worth to stress that in reality the blue luminosity 
reflects the stellar mass only in galaxies with no active star forming regions, whereas in  
late types it is the result of the combined effects of the emission by both 
old stars and young stellar population, along with the absorption by dust grains.\\  
By looking at the type Ia rate as a function of the galactic type, 
we reproduce the observed data for E/S0 and Irr galaxies. 
In the case of S0a/b and Sbc/d galaxies, the data are overestimated and underestimated, 
respectively. 
Our predictions indicate that S0a/b galaxies have a higher type Ia SNR per unit luminosity    
than Sbc/d galaxies because of their higher number of tye Ia SN progenitors, due to 
a past average star formation rate higher than Sbc/d galaxies. \\
Concerning SN Ib/c, according to our predictions the rate in E/S0 is zero, since in these galaxies star 
formation stopped more than 12  Gyr ago. 
For the S0a/b  and Sbc/d galaxies,  with a Scalo IMF (open circles in Fig.~\ref{SNu_})  our models underestimate the observed rates. 
For Irr galaxies, the predicted Ib/c rate lies in the lower part of the 
error bar. 
This fact probably indicates that we are underestimating the Ib/c rates, i.e. that 
the  single stars with masses $M> 25 M_{\odot}$ are not sufficient to explain the observed type Ib/c SN rates and that other progenitors 
should be taken into account. 
 On the other hand, we note that with the assumption of a Salpeter IMF (inverted triangles in Fig.~\ref{SNu_}) in spiral discs, the type Ib/c SN rates are well accounted for.  \\
Finally, we note that the predicted type II SN rates are in very good agreement with the data.  
Given the many sources of uncertainty in determining the blue luminosity of 
galaxies (mostly dust extinction  and inclination effects) as well as the galactic Hubble types, we find that 
our theoretical picture provides a good fit of the observed rates. \\

\subsection{SN rates per unit mass}

The K band luminosity is dominated by old stellar populations, which contribute to 
the bulk of the stellar mass of present-day galaxies. 
For this reason, the K luminosity represents a better tracer of the stellar mass 
than the B band luminosity. Moreover, the   
K band luminosity is not affected by dust extinction effects, 
hence its determination is in principle more robust than the one for  the B band. 
The K band luminosity  has been used by Mannucci et al. (2005) to derive the SN rate per unit mass. 
In Figure~\ref{SNuM_}, we show the observed and predicted present-day  
Ia (lower panel), Ib/c (middle panel) and II (higher panel)  
SN rate per unit mass as a function of the galactic morphological type. 
Also in this case, the observations show an increase in the rates 
from early types to late types. In general, this effect is even more pronounced
than in Figure~\ref{SNu_}. We notice that this effect is accounted for  
by our predictions. Furthermore, the agreement between the model predictions and the observations 
here is better than in Figure~\ref{SNu_}. \\
The type Ia SNR per unit mass is reproduced very well for all galaxy types. 
The discrepancy between the observed and predicted Ib/c rates for S0a/b, Sbc/d  (with a Scalo IMF) and Irr is still present. 
This confirms that, if the IMF in spiral disks is the one of Scalo (1986), 
massive WR stars are unlikely to represent the only progenitors of type Ib/c SN. 
On the other hand, it is important to note that the assumption of a Salpeter IMF (inverted triangles in Fig.~\ref{SNuM_}) in S0a/b and Sbc/d galaxies, 
allows us to reproduce the observed type Ib/c SN rates. 

\subsection{SN type Ib/c rates from binary systems}

Other possible progenitors for type Ib/c SNe can be less massive stars (i.e. stars with initial masses 
12-20  $M_{\odot}$)  in close binary systems, ending their lives as He stars and ejecting their 
H envelope by means of mass transfer (Filippenko 1991) and finally exploding as core collapse SNe. 
Therefore, we recalculate the type Ib/c and II SN rates by 
assuming that the Ib/c SN progenitors are both single WR stars and close massive binaries. 
In this case, the  uncertain parameter is the proportion of massive stars in close binary systems which will give rise to Type Ib/c SNe, the equivalent of the parameter $A_{Ia}$ described in Sect. 2.6. 
This quantity is practically unknown, as it is $A_{Ia}$, and it should be 
determined as the best value to reproduce the present-day type Ib/c SNRs 
in our Galaxy.  
However, we can attempt to give a realistic estimate of this quantity, by  
assuming that in our Galaxy, half of the massive stars are in binary systems.  
We assume a close binary frequency of $30 \%$, i.e. similar to the close binary 
frequency  predicted for low mass systems (Jeffries \& Maxted 2005). In this way,  
we tentatively estimate, for the Scalo (1986) IMF that a fraction:
\begin{equation}
 A_{Ib/c, Sc} = 0.5 \cdot 0.3 = 0.15 
\end{equation}
of the massive stars with masses in the range 12-20 $M_{\odot}$ may represent  
possible type Ib/c SN progenitors.We remind here that, like for type Ia SNe,  we do not consider the existence of massive binaries other than for the progenitors of type Ib/c SNe, and treat all the stars, binaries and single, as single stars from the point of view of the IMF.  

As in the case of the fraction $A_{Ia}$ of binary 
systems which can end as type Ia (see eq. 6), the quantity $A_{Ib/c}$ depends on the assumed stellar IMF. 
Here, we assume that: 
\begin{equation}
 \frac{A_{Ib/c,Salp}}{A_{Ib/c, Sc}} = \frac{A_{Ia,Salp}}{A_{Ia, Sc}} 
\end{equation}
where $A_{Ia,Salp}$ and $A_{Ia, Sc}$ refer to the Salpeter and Scalo IMF, respectively, and are chosen according to Matteucci et al. (2003). 
By subtracting  the quantity 
\begin{equation} 
  R_{Ib}^{MB}(T_{0}) = A_{Ib/c} \cdot  R_{II}(T_{0})    
\end{equation}
(where $A_{Ib/c}=A_{Ib/c, Salp}$ for E/S0 and Irr and $A_{Ib/c}=A_{Ib/c, Sc}$ for S0a/b and Sbc/d, respectively) 
from the predicted type II SN rate calculated at the present time ($R_{II}(T_{0}) $), and by adding 
this same quantity to the previously calculated current type Ib/c SN rate 
($R_{Ib} (T_{0})$), we obtain:

\begin{equation}
  R_{II}'(T_{0}) = R_{II} - R_{Ib}^{MB}(T_{0})   
\end{equation}

\begin{equation}
  R_{Ib}'(T_{0}) = R_{Ib}(T_{0}) + R_{Ib}^{MB}(T_{0})    
\end{equation}
These quantities are the 
revised calculations for the type II and  Ib/c SN rates, indicated by 
$R_{II}'(T_{0})$ and $R_{Ib}'(T_{0})$, respectively.  
In Figure~\ref{SNuM_} , these new values are indicated as dotted circles. 
We note that the newly calculated type Ib/c SN rates are in very good agreement with the observations. 
Our conclusions are that single WR stars, exploding as type Ib/c SNe, are not sufficient to account for 
the observed  Ib/c SNR per 
unit mass. Less massive stars, with initial masses $12\le M/M_{\odot} \le 20$,   
in close  binary systems, give a significant contribution to the Ib/c SN rates. 
 It is important to stress that, for S0a/b and Sbc/d galaxies, 
this conclusion holds only in the case of the assumption of a Scalo (1986) IMF.  \\
In Table~\ref{rates}, we present all the predicted present-day  estimates for the type Ia, Ib/c and II SN rates in the 
four morphological types and we compare them with the available observational estimates. 
The SN rates are calculated per unit B band luminosity and per unit mass. 
Hubble types are reported in the first column. The observed and predicted values for the type Ia, Ib/c and II 
SN rates are reported in the second, third and fourth double columns, respectively. 
For S0a/b, Sb/c in the case of a Scalo IMF and Irr galaxies, 
the entries for the predicted type Ib/c SN rates in Table~\ref{rates} include also the contribution by massive binaries and have been calculated 
according to equation (16).  For S0a/b and Sb/c, in the case of the adoption of  Salpeter IMF, the entries include only the contribution 
by single stars with masses $M \ge 25 M_{\odot}$.
In general, the agreement between the model predictions and the data is rather good. 
Given the several sources of uncertainty in recognizing  the 
galactic Hubble types and in some cases the SN types, we find that our models provide a good fit of the observed rates. \\

\renewcommand{\baselinestretch}{1.0}
\begin{table*}
\centering
\caption{ Observed and predicted SNR per unit B luminosity and per unit mass. 
}
\begin{tabular}{lcccccccc}
\\[-2.0ex] 
\hline
\multicolumn{9}{c}{SN rate per unit B  luminosity (SNu)}\\
\hline
\\[-2.5ex]
\multicolumn{1}{l}{Hubble Type}&\multicolumn{2}{c}{Ia}&\multicolumn{1}{c}{}&\multicolumn{2}{c}{Ib/c}&\multicolumn{1}{c}{}&\multicolumn{2}{c}{II}\\
\multicolumn{1}{c}{}&\multicolumn{1}{c}{Obs$^1$}&\multicolumn{1}{c}{Pred}&\multicolumn{1}{c}{}&\multicolumn{1}{c}{Obs$^1$}&\multicolumn{1}{c}{Pred}&\multicolumn{1}{c}{}&\multicolumn{1}{c}{Obs$^1$}&\multicolumn{1}{c}{Pred}\\

\hline
\hline
\\[-1.0ex]
E/S0                 &   $0.17\pm0.06$     &   0.19  &       & $<0.01$          &  0.     &   & $<0.02$       &  0.      \\
\vspace*{-2.5mm} \\
\hline
\vspace*{-2.5mm} \\
S0a/b                &   $0.17\pm0.07$     &   0.3    &       & $0.10\pm0.06$    &  0.06   &   & $0.40\pm0.18$ &  0.28    \\
(Scalo IMF)          &                     &          &       &                  &         &   &               &          \\ 
S0a/b                &   $0.17\pm0.07$     &   0.29   &       & $0.10\pm0.06$    &  0.1    &   & $0.40\pm0.18$ &  0.76    \\
(Salpeter IMF)       &                     &          &       &                  &         &   &               &          \\ 
\vspace*{-2.5mm} \\
\hline
\vspace*{-2.5mm} \\
Sbc/d                &   $0.20\pm0.08$     &   0.12   &       & $0.13\pm0.07$    &  0.09   &   & $0.81\pm0.33$ &  0.50    \\
(Scalo IMF)          &                     &          &       &                  &         &   &               &          \\ 
Sbc/d                &   $0.20\pm0.08$     &   0.09   &       & $0.13\pm0.07$    &  0.15   &   & $0.81\pm0.33$ &  1.1     \\
(Salpeter IMF)       &                     &          &       &                  &         &   &               &          \\ 
\vspace*{-2.5mm} \\
\hline
\vspace*{-2.5mm} \\
Irr                  &   $0.38\pm0.15$     &   0.5   &       & $0.21\pm0.15$    &  0.26   &   & $0.62\pm0.37$ &  0.48    \\
\vspace*{-2.5mm} \\
\hline
\multicolumn{9}{c}{SN rate per unit mass (SNuM)}\\
\hline
\\[-2.5ex]
\multicolumn{1}{l}{Hubble Type}&\multicolumn{2}{c}{Ia}&\multicolumn{1}{c}{}&\multicolumn{2}{c}{Ib/c}&\multicolumn{1}{c}{}&\multicolumn{2}{c}{II}\\
\multicolumn{1}{c}{}&\multicolumn{1}{c}{Obs$^2$}&\multicolumn{1}{c}{Pred}&\multicolumn{1}{c}{}&\multicolumn{1}{c}{Obs$^2$}&\multicolumn{1}{c}{Pred}&\multicolumn{1}{c}{}&\multicolumn{1}{c}{Obs$^2$}&\multicolumn{1}{c}{Pred}\\

\hline
\hline
\\[-1.0ex]
E/S0                 & 0.042$_{-0.013}^{+0.015}$    & 0.05  &       & $<$0.009                   &  0.     &   & $<$0.012                    &  0.     \\
\vspace*{-2.5mm} \\
\hline
\vspace*{-2.5mm} \\
S0a/b                & 0.062$_{-0.024}^{+0.026}$    & 0.10 &       & 0.034$_{-0.017}^{+0.025}$   &  0.02   &   & 0.11$_{-0.051}^{+0.056}$    &  0.12   \\
(Scalo IMF)          &                     &        &       &                  &         &   &               &          \\ 
S0a/b                & 0.062$_{-0.024}^{+0.026}$    & 0.06  &       & 0.034$_{-0.017}^{+0.025}$   &  0.02   &   & 0.11$_{-0.051}^{+0.056}$   &  0.16   \\
(Salpeter IMF)       &                     &          &       &                  &         &   &               &          \\ 
\vspace*{-2.5mm} \\
\hline
\vspace*{-2.5mm} \\
Sbc/d                & 0.16$_{-0.06}^{+0.064}$     & 0.11   &       & 0.11$_{-0.056}^{+0.070}$    &  0.08   &   & 0.70$_{-0.28}^{+0.29}$      &  0.44    \\
(Scalo IMF)          &                             &        &       &                             &         &   &                             &          \\ 
Sbc/d                & 0.16$_{-0.06}^{+0.064}$     & 0.08   &       & 0.11$_{-0.056}^{+0.070}$    &  0.13   &   & 0.70$_{-0.28}^{+0.29}$      & 0.96     \\
(Salpeter IMF)       &                     &          &       &                  &         &   &               &          \\ 
\vspace*{-2.5mm} \\
\hline
\vspace*{-2.5mm} \\
Irr                  & 0.73$_{-0.29}^{+0.40}$       & 0.82   &       & 0.51$_{-0.36}^{+0.63}$      &  0.44   &   & 1.6$_{-0.95}^{+1.33}$         &  0.80    \\
\vspace*{-2.5mm} \\
\hline
\hline
\end{tabular}
\label{rates}
\flushleft
Observed values: $^1$  Cappellaro et al. (1999); $^2$ Mannucci et al. 2005.\\
\end{table*}

\begin{figure*}
\centerline{\includegraphics[height=19pc,width=19pc]{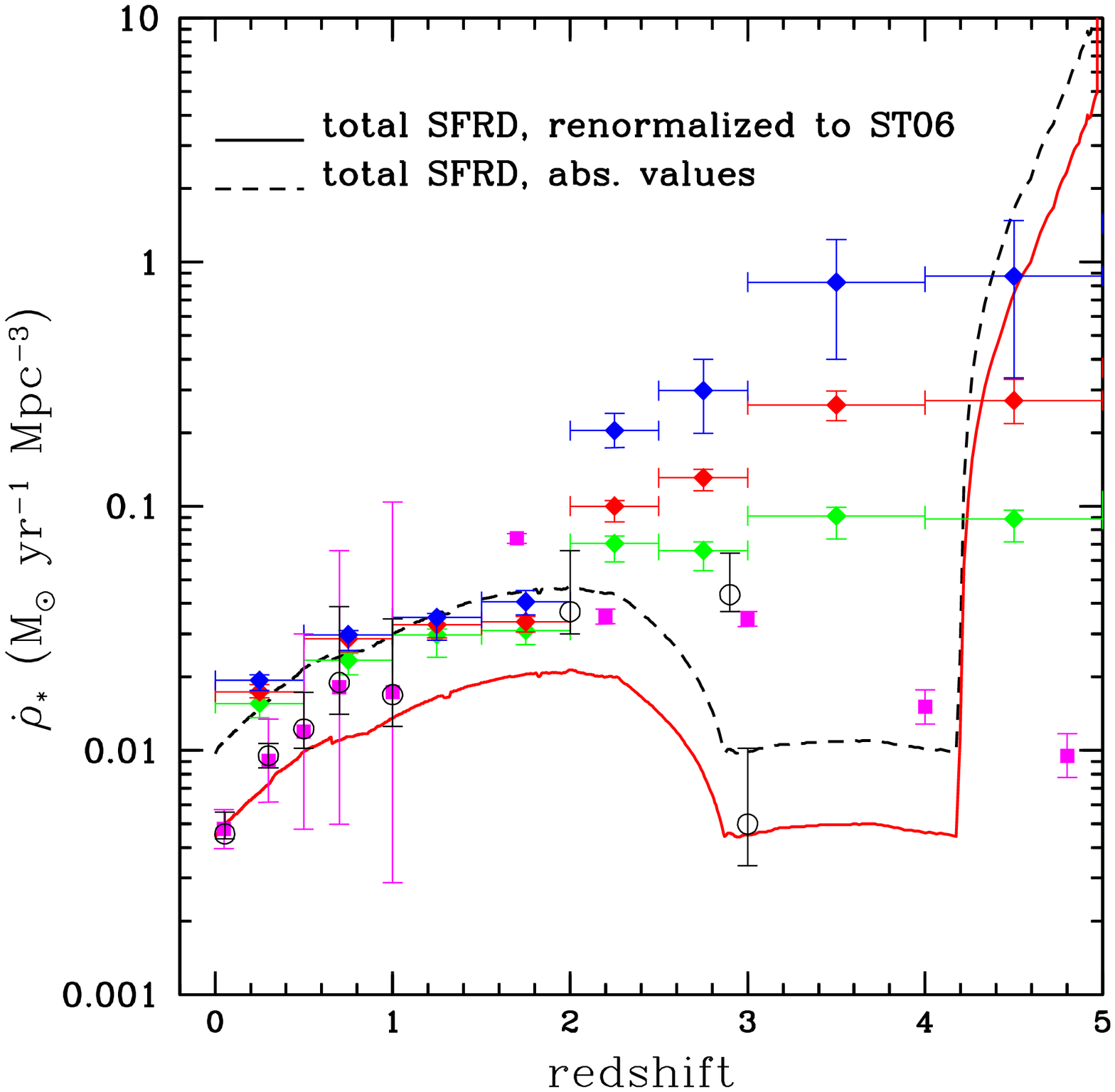}} 
\caption[]{
Predicted and observed star formation rate density assuming that galaxy formation started 
 at $z_{f}=5$. The dashed line represents the predicted 
total comoving SFR density, given by the sum of the contributions of all the different morphological types. 
The solid line is the predicted SFR density, re-normalized to the $z=0.05$ value by Sawicki \& Thompson (2006). 
The solid squares are the observational data by Sawicki \& Thompson (2006). The open circles are the data by Schiminovich et al. (2005).  
The solid diamonds are taken from Lanzetta et al. (2002). 
}
\label{SFRD}
\end{figure*}

\subsection{The Evolution of the cosmic star formation rate density} 

A significant test for any galaxy evolution model is represented by the study of the cosmic star formation history 
(Hopkins 2004, Sawicki \& Thompson 2006). 
In this paper, we calculate the star formation rate density  $\dot{\rho_{*}}(z)$   as indicated in Calura \& Matteucci (2003):
\begin{equation}
\dot{\rho_{*}}(z) = \sum_{i} \rho_{B i}(z) \, (\frac{M}{L})_{B i}(z) \, SFR_{m i}(z)
\end{equation}  
where $ \rho_{B i}$, $(\frac{M}{L})_{B i}$ and $SFR_{m i}(z)$ are the B luminosity density, the B mass-to-light ratio and 
star formation rate per unit mass for the galaxies of the $i$-th morphological type, respectively. 
For purposes of comparison with all the observational data, hereinafter we 
adopt a Lambda-cold dark matter cosmology
($\Lambda$CDM, $\Omega_{0}=0.3$, 
 $\Omega_{\Lambda}=0.7$) and $h=0.73$. For all galaxies, we assume that the star formation started at redshift $z_{f}=5$. 
According to this cosmological model, the redshift $z_{f}=5$ corresponds to a 
lookback time of 11.8 Gyr.\\
The points in Figure~\ref{SFRD} represent estimates based on observations in the UV band, which is 
contaminated by  dust extinction effects.  
However, since the extent of the attenuation by dust is highly uncertain (see Steidel 
et al. 1999, Hopkins 2004), we have chosen to plot the data uncorrected for
dust extinction. 
 The dashed line in  Figure~\ref{SFRD} represents 
the predicted evolution  of the cosmic star formation history, obtained by means of our models. 
The solid line represents 
the predicted cosmic SFR, renormalized to the value calculated at $z=0.05$ by Sawicki \& Thompson (2006). 
The solid line is drawn in order to stress that the evolution of the cosmic SFR observed between redshift $z=1$ and $z=0$ 
is well reproduced by our models. 
The shape of the  SFR density at redshifts $z>2$ is highly uncertain (Hopkins 2004). 
The fit to the points can be improved by shifting the redshift of formation $z_{f}=5$  towards higher values. The peak predicted by 
means of our models is due to 
the strong star formation in the progenitors of E/S0 galaxies and might not be observable if the
associated starbursts occurred in sites heavily obscured by dust. In fact, in their study of the UV luminosity density evolution, 
Calura, Matteucci \& Menci (2004) have shown that, once dust obscuration effects are 
taken into account, the predicted peak at $z=5$ of the UV luminosity density,  due to early type galaxies, levels off and their 
predictions become consistent with the observed values.

\subsection{The Evolution of the supernova rate density}

Once we have verified that our galaxy evolution picture reproduces the observed cosmic star 
formation history, now we focus on the cosmic SNR density and its evolution. 
The SNR density (SNRD)  is defined as the SNR per unit cosmic volume. 
This quantity is important since it gives information on how the total galaxy populations   
evolve with cosmic time. This study is also useful to understand the contributions 
that the different galactic morphological types give to the total 
supernova rate and, as a consequence, to the global rate of metal production. 
The SNRD values observed so far concern type Ia SNe and 
core collapse (CC) SNe, which include the categories of type Ib/c and II. For this reason,  
from this moment on 
we ignore the distinction between type Ib/c and II SNe. The CC SNRD is given by the sum of the type Ib/c and II SNRDs.

\begin{figure*}
\centerline{\includegraphics[height=19pc,width=19pc]{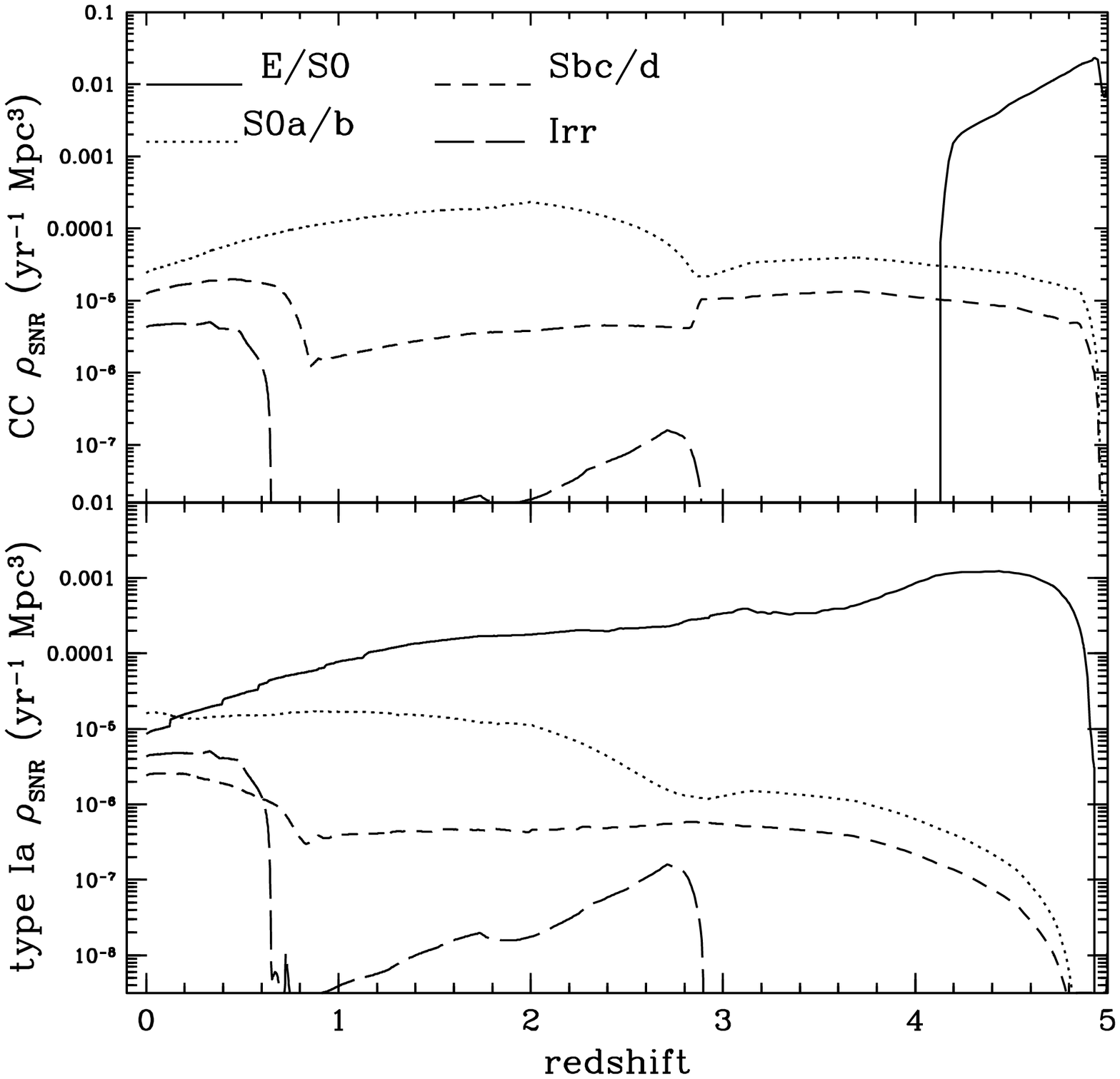}} 
\caption[]{Predicted type Ia (lower panel) and core collapse (higher panel) 
SN rate density as a function of redshift for various galactic morphological types. 
Solid lines: contributions by E/S0 galaxies. Dotted lines: contributions by S0a/b. 
Short-dashed lines: contributions by Sbc/d. Long-dashed lines: contributions by Irr galaxies. 
}
\label{SNRD}
\end{figure*}

\begin{figure*}
\leftline{\includegraphics[height=19pc,width=23pc]{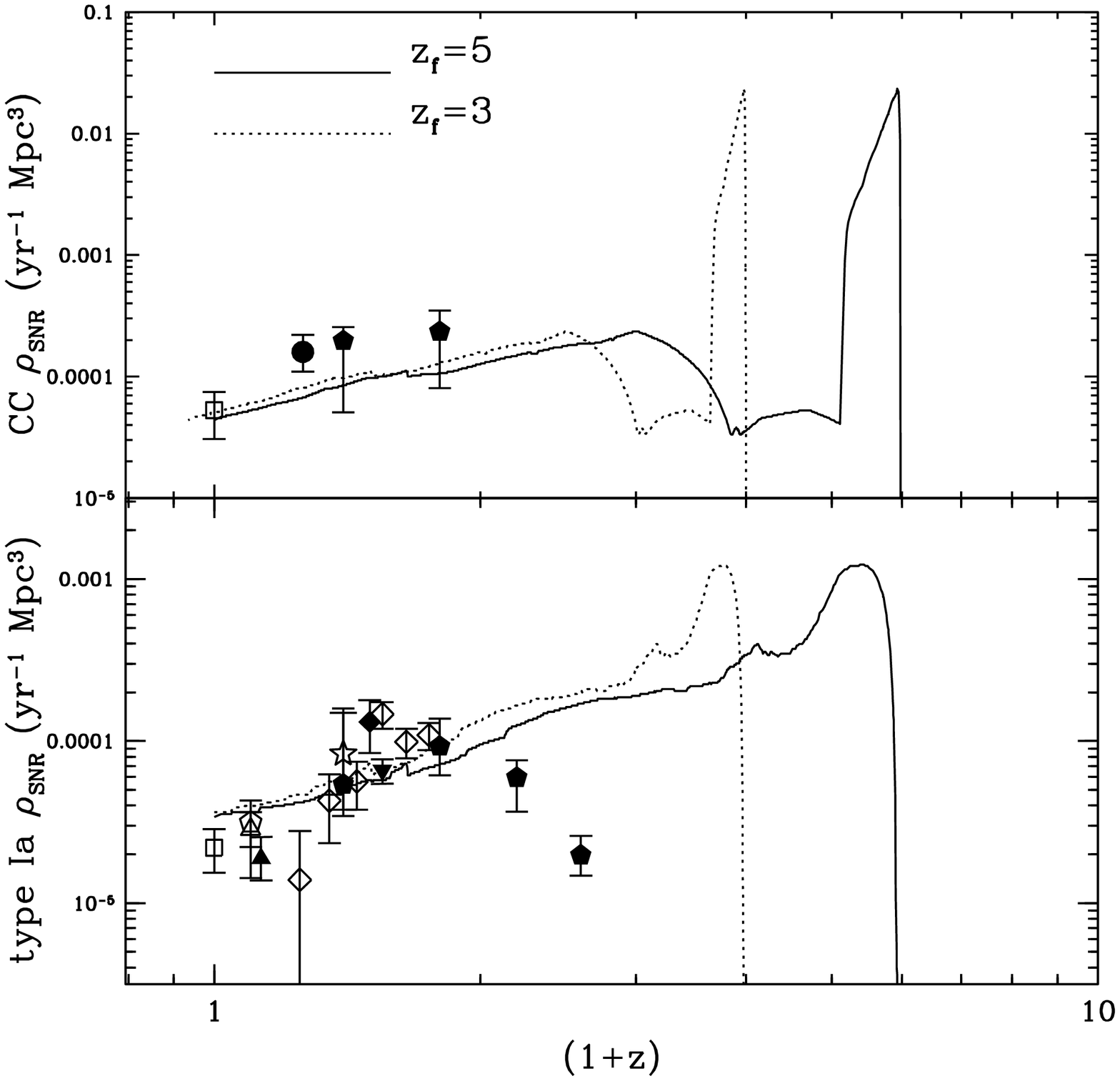}
\leftline{\includegraphics[height=19pc,width=19pc]{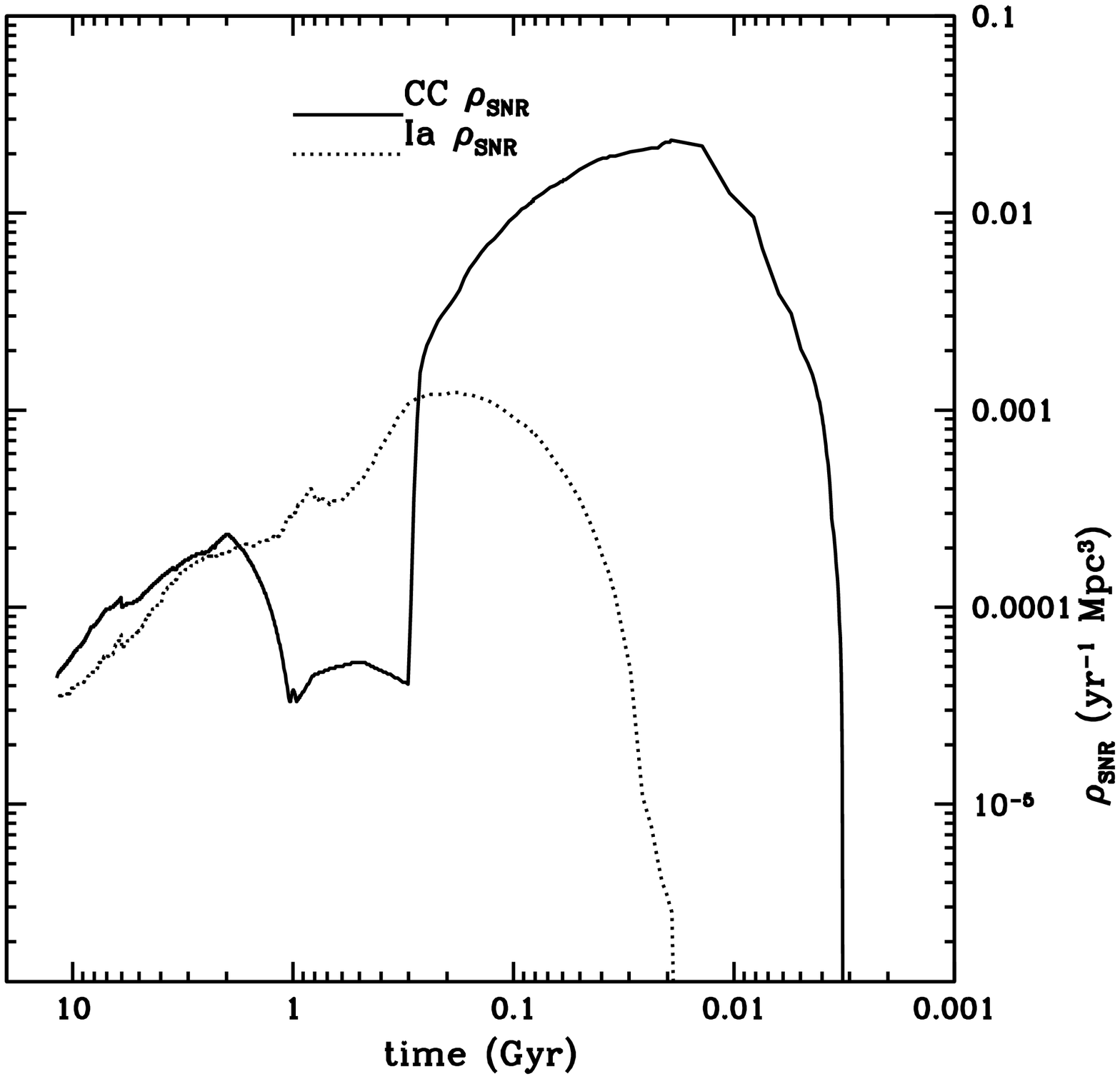}} } 

\caption[]{
Observed and predicted type Ia (left lower panel) and core collapse (left higher panel) cosmic 
SN rate density as a function of redshift and as a function of cosmic time (right panel). The predictions have been calculated by assuming two possible 
redshifts of formation for all galaxies: $z_{f}=3$ (dotted lines) and  $z_{f}=5$ (solid lines). 
Open squares: Cappellaro et al. (1999). Open triangle: Madgwick et al. (2003). Open pentagon: Hardin et al. (2000). 
Solid diamond: Tonry et al. (2003). Solid triangle: Blanc et al. (2004). Inverted solid triangle: Pain et al. (2002). 
Open star: Pain et al. (1996). Open diamonds: Barris et al. (2006). Solid pentagons: Dahlen et al. (2004). 
Solid circle: Cappellaro et al. (2005). 
}
\label{SNRDdata}
\end{figure*}

\begin{figure*}
\centerline{\includegraphics[height=19pc,width=19pc]{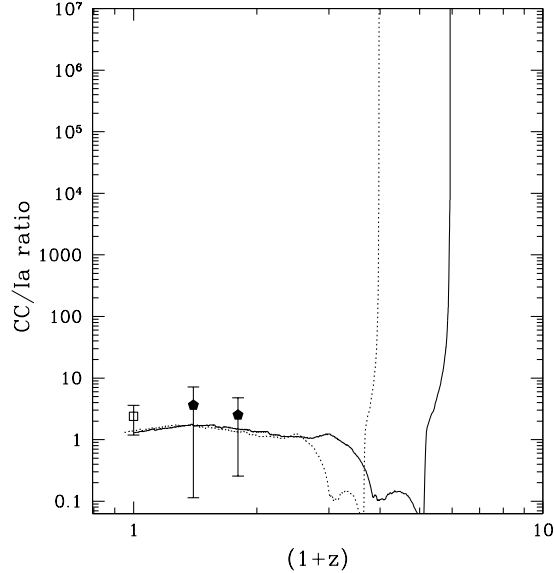}} 
\caption[]{
Observed and predicted Core Collapse to type Ia  SN rate ratio as a function of redshift. 
The predictions have been calculated by assuming two possible 
redshifts of formation for all galaxies: $z_{f}=3$ (dotted lines) and  $z_{f}=5$ (solid lines). 
Open squares: Cappellaro et al. (1999). Solid pentagons: Dahlen et al. (2004). 
}
\label{SNRratio}
\end{figure*}

For the $k-$th galactic morphological type, the type $\gamma$  SNRD (expressed as $SN \, yr^{-1} \, Mpc^{-3}$)
as a function of the redshift $z$ is : 
\begin{equation}
 \rho_{SNR, \gamma}(z) = \rho_{B, k}(z)  \, SNu_{k}^{\gamma} \, 100^{-1} \, yr^{-1} 10^{-10} L_{B, \odot}
\label{snrdeq}
\end{equation} 
$ \rho_{B, k} (z)$ is the B band luminosity density (BLD) for the $k-$ the morphological type. 
At $z=0$ the BLDs for the single galaxy types are the ones observed by Marzke et al. (1998), 
who determined the local B band luminosity function for three morphological types: 
early-types, spirals   and irregular galaxies. 
We assume that 
the early type BLD represents the contribution by E/S0 galaxies.  
The morphological fractions for the S0a/b and Sbc/d galaxies are the ones observed 
by Nakamura et al. (2003), who have shown that the stellar mass density of spiral disks 
is dominated by S0a/b galaxies, with a 73\% contribution. The remainder 
27\% is the contribution by Sbc/d galaxies. 
These fractions, multiplied by the total 
spiral BLD as observed by Marzke et al. (1998), give the local S0a/b and Sbc/d BLDs. \\ 
At a generic redshift $z$, the luminosity density is calculated as described in Calura \& Matteucci (2003, 2004). 
The quantity $SNu_{k}^{\gamma}$ is the SNR, expressed in SNu, for the $k-$th galactic type 
and for the $\gamma$-th SN type. 
In Figure~\ref{SNRD} we show the predicted redshift evolution of the  type Ia SNRD (lower panel) and CC SNRD 
(upper panel) for E/S0 galaxies (dashed lines), S0a/b (solid lines), Sbc/d (dash-dotted lines) 
and irregular galaxies (dotted lines). 
We note that the type Ia SNRD (Figure~\ref{SNRD}, lower panel) is dominated by 
E/S0 galaxies throughout most of the cosmic time. 
S0a/b are the second main contributors to the type Ia SNRD at any redshift. Sbc/d and Irr give a minor contribution throughout 
most of the cosmic time. At the present time, the type Ia SNRD is dominated by S0a/b galaxies.\\
The CC SNRD (Figure~\ref{SNRD}, upper panel) at early epochs is dominated by 
the progenitors of E/S0 galaxies, which, as predicted by Calura \& Matteucci (2003),   
cause also a peak in the cosmic SFR density. The S0a/b and Sbc/d galaxies dominate the CC SNRD throughout most of the cosmic time.  
At the present time, the CC SNRD is dominated by S0a/b and Sbc/d galaxies, with Irr giving a minor contibution. 
In E/S0 galaxies, the present-day CC SNRD is zero.\\
In Figure~\ref{SNRDdata}, we show the predicted and observed 
redshift evolution of the total type Ia (left lower panel) and CC (left upper panel) SNRD, along with the time 
evolution of the Ia and CC SNRD (right panel). 
By looking at the time evolution of the type Ia and CC SNRD (Figure~\ref{SNRDdata}, right panel), we note that 
the type Ia and CC SNRD present a peak at $0.2$ Gyr  and  $0.02$ Gyr  
after the beginning of the star formation, respectively. 
The delay between these two peaks is due to the different mass ranges, and consequently explosion timescales, of type Ia and CC SNe. \\
In the left panels, the predictions have been calculated by assuming two different redshifts of galaxy 
formation: $z_{f}=3$ (dotted lines) and  $z_{f}=5$ (solid lines). 
The sources of the observed SNRD are reported in the caption of Figure~\ref{SNRDdata}. 
All the observed SNRDs have been normalized to the local BLD used in this work, i.e. the one by 
Marzke et al. (1998). 
Concerning the type Ia SNRD, the majority of the observed data up to redshift $z \sim 1$ 
(i.e. $(1+z) \sim 2$ in Fig.~\ref{SNRDdata}) 
are well reproduced by our models. 
On the other hand, the points at redshift $z>1$ (corresponding to $(1+z)>2$ in Figure~\ref{SNRDdata}) 
observed by Dahlen et al. (2004) are overestimated by our predictions. 
These data are very difficult to reproduce by means of standard SN rate models (Dahlen et al. 2004; Mannucci et al. 2005), unless a  
significant delay time ($\tau_{d} = 4$ Gyr) between the epoch of star formation and the explosion of type Ia SN is assumed. 
In principle, this means that 
one should wait a time comparable to $\sim$ 4 Gyr after 
 the beginning of star formation, in order 
to observe a significant contribution from type Ia SN to the chemical enrichment of any astrophysical 
system. 
This conclusion is in strong contrast with the results of any chemical evolution model describing the Milky Way galaxy. 
It is well known that CC SNe enrich the galactic ISM of $\alpha$ elements, such as O and Mg. 
On the other hand, type Ia SNe are the main  producers of Fe-peak elements. 
For this reason, the study of the ($\alpha$/Fe) ratio in astrophysical objects carries 
important information on the star formation history and on the age of the systems. 
Chemical evolution models indicate that in the solar neighbourhood, the time at which 
the Fe production from SNe Ia starts to become important is $\sim 1$ Gyr after the beginning of star formation  
(Matteucci \& Greggio 1986, Matteucci \& Recchi 2001). This timescale allows us to explain the ($\alpha$/Fe) values as a function of the (Fe/H) ratio
observed in Galactic field stars.\\ 
A delay time of $\sim 4$ Gyr is 
also in contrast with 
chemical evolution studies of Damped Lyman Alpha (DLA) systems. 
DLAs are Quasar absorbing systems characterized by 
high neutral gas content (with typical values for the neutral H column density of $N(HI) \ge 2 \cdot 10^{20} cm^{-2}$)  
and 
metal abundances  which can span from $\sim 1/100$ solar up to 
the solar value (Wolfe, Gawiser \& Prochaska 2005).  
Recently, by  comparing the abundance ratios observed in a sample of DLAs 
with the predictions from chemical evolution models, it has been possible to determine for the first time 
the age of these systems (Dessauges-Zavadsky et al. 2004), with values ranging from  $\sim 0.1$ Gyr up to $1-2$ Gyr. 
In general, DLAs show solar (O/Fe) ratios (Calura, Matteucci \& Vladilo 2003). This means that in these systems, 
the contribution of type Ia SNe to the chemical enrichment of the ISM has already been significant. 
Given their ages, 
with a typical timescale of $\sim 4$ Gyr for type Ia SNe, it would be impossible to find solar (O/Fe) values in DLAs. 
Moreover, recently Mannucci et al. (2006) have shown that a large fraction of SNe Ia should arise from fast systems, 
i.e. exploding on timescales of the order of 40 Myr, to explain the type Ia SN rates observed in radio loud ellipticals. 
For these reasons, we believe that the decline in the cosmic Ia  SNRD as observed by Dahlen et al. (2004) should be 
regarded with caution. 
These data are the first collected at redshift $>1$ for the type Ia SN rate, and are likely to represent 
lower limits to the actual values. 
Our suggestion is that, before drawing definitive conclusions on the behaviour of the type Ia SNRD at redshift $z>1$, 
we have to await for more data from future surveys.\\ 
Finally, from the lower panel of Fig.~\ref{SNRDdata}, we note that the assumption that galaxy formation started 
at $z_{f}=5$ or at $z_{f}=3$ has a minor impact on the theoretical results.\\
In the upper panel of Figure~\ref{SNRDdata}, we show the predicted and observed evolution of the CC SNRD. 
In this case, the available data are scant. The available measures have been derived only by Cappellaro et al. (1999), Dahlen et al. 
(2004) and Cappellaro et al. (2005). \\
The observed local CC SNRD is reproduced by our models. On the other hand, all the SNRD values observed 
at redshift $z\ge 0.3$ are underestimated by our results, although consistent with the error bars. 
On the other hand, our models allow us to reproduce the observed evolution of the cosmic star 
formation, which is a quantity proportional to the CC SN rate density. We believe that, owing to paucity of 
the observational data, no firm conclusion can be drawn on the evolution of the CC SN rate density at redshift 
$z>0$. \\
In Figure~\ref{SNRratio}, we show the redshift evolution of the predicted CC/Ia SN rate ratio, along with the 
few available data which have been collected for both quantities (Cappellaro et al. 1999, Dahlen et al. 2004). 
The predictions indicate that the CC/Ia ratio does not vary significantly between $z=0$ and $z=1$. 
The predictions are all consistent with the available observations. 
At higher redshift, according to our results the CC/Ia ratio should show a large peak, in correspondence to the epoch of 
major spheroid formation. This effect could be in principle observable in the future, thanks to 
the next generation space and ground-based telescopes. 

\section{Conclusions}
In this paper, we have studied the type Ia, Ib/c and II SN rates for galaxies of different morphological types. 
We have built four different chemical evolution models, each one representing a different Hubble type: 
E/S0, S0a/b, Sbc/d and Irr galaxies, respectively. We have interpreted the Hubble sequence as due to a decreasing star 
formation efficiency and an increasing infall timescale going from early to late type galaxies.
We have then used these models to study the evolution of the SN rates per unit luminosity and per unit mass as a function 
of time and as a function of the Hubble type. This kind of study 
can provide useful constraints on the possible SN progenitor models and on several galaxy evolution parameters. 
We have compared the observed local SN rates with our predictions, finding a generally good agreement. 
Finally, we have investigated the redshift evolution of the core collapse and type Ia SN rate density, a quantity 
very important to understand the global evolution of star formation in the universe. 
Our results can be summarized as follows:\\
1) The local observations indicate an increase of both the SN rate per unit luminosity and mass  from early to late galaxy 
types. This feature is well reproduced by our models.  
According to our results, this effect is due to the fact that 
the latest Hubble types have in general the highest star formation rates per unit mass.\\
2)  If we adopt a Scalo (1986) IMF in S0a/b and Sbc/d galaxies, our results show that 
massive (i.e. with initial masses $M>25 M_{\odot}$) single W-R stars, losing the H envelope by means of stellar winds and exploding 
as type Ib/c SNe, are not sufficient to account for 
the observed  Ib/c SN rate per unit mass.  In this case, 
less massive stars, i.e. with initial masses $12 \le M/M_{\odot} \le 20$, 
in close binary systems give a significant contribution to the local Ib/c SN rates. On the other hand, by adopting a Salpeter (1955) 
IMF for all morphological types, it is possible to explain the observed type Ib/c SN rates without 
resorting to the contribution by massive stars in close binary systems. However, Romano et al. (2004) have studied the effects 
of varying the IMFs in the chemical evolution models of the Milky Way galaxy. These authors have found that the Salpeter IMF leads to a serious overestimation  
of the observed solar abundances, showing that, for the modelling of Milky Way disk, the Scalo (1986) IMF is favoured over the Salpeter (1955) one.\\
3) The main contributors to the type Ia SN rate density throughout most of the 
cosmic time are E/S0 galaxies. \\
During the epoch of E/S0 formation, the progenitors of these galaxies dominate the total 
core collapse SN rate density. S0a/b and Sbc/d galaxies dominate the core collapse SN rate density throughout the 
remainder of the cosmic time, with Irr galaxies giving 
a minor contribution. \\
4) The predicted type Ia cosmic  SN 
rate density increases with redshift, reaching a peak at redshift $z \ge3$, which is due to the contribution by massive spheroids. 
Our models allow us to reproduce the observed type Ia SN rate density up to redshift $z\sim 1$. 
At higher redshifts, our predictions overestimate the few available data, which  
cannot be reproduced unless a  significant delay time ($\tau_{d} = 4$ Gyr) between the epoch of star formation and 
the explosion of type Ia SNe is assumed, as shown by Dahlen et al. (2004). 
This  delay time  is in strong contrast with chemical evolution results 
from studies of the ($\alpha$/Fe) ratio in the Milky Way and in Damped Lyman Alpha systems. 
It is also in contrast with the empirical type Ia SN rate recently derived by Mannucci et al. (2006), which agrees with 
our assumptions on the 
type Ia SN rate, which includes systems exploding only after $\sim 0.03$ Gyr from the beginning of star formation.
More data on the SN Ia at high redshift are necessary before drawing firm conclusions. \\
5) At $z=0$, we reproduce the observed CC SN rate density. At redshift  $z>0$, 
the few observations of the CC  SN rate density are underestimated by our results, although consistent with the error bars. 
To draw firm conclusions on the behaviour of the CC SN rate at redshift $z>0$, more observational data are needed. \\
6) Our predictions indicate that the ratio of the core collapse to Ia SN rate 
should present a peak in 
correspondence of the major spheroid formation epoch. In the future, the next generation telescopes could allow us to observe 
this effect. 

 \acknowledgments
We are grateful to Raul Jimenez for having provided us with the 
software necessary to calculate the spectra of the simple stellar populations.  

 \acknowledgments

\end{document}